A Systematic Review on Crimes facilitated by Consumer Internet of Things Devices

Ashley M. Brown
Department of Security and Crime Science, University College London (UCL),
ashley.brown.21@ucl.ac.uk

Nilufer Tuptuk
Department of Security and Crime Science, University College London (UCL),
n.tuptuk@ucl.ac.uk

Enrico Mariconti
Department of Security and Crime Science, University College London (UCL),
e.mariconti@ucl.ac.uk

Shane D. Johnson
Department of Security and Crime Science, University College London (UCL),
shane.johnson@ucl.ac.uk

**Abstract:** It is well documented that criminals use IoT devices to facilitate crimes. The review process follows a systematic approach with a clear search strategy, and study selection strategy. The review included a total of 543 articles and the findings from these articles were synthesised through thematic analysis. Identified security attacks targeting consumer IoT devices include man-in-the-middle (MiTM) attacks, synchronisation attacks, Denial-of-Service (DoS), DNS poisoning and malware, alongside device-specific vulnerabilities. Besides security attacks, this review discusses mitigations. Furthermore, the literature also covers crime threat scenarios arising from these attacks, such as, fraud, identity theft, crypto jacking and domestic abuse.

## 1. INTRODUCTION

Many electronic devices are now Internet-connected and capable of interacting with each other. Such devices are collectively referred to as the Internet of Things (IoT). Globally, the total number of IoT connections is estimated to reach 31 million in 2030, up from 14 million in 2024 [1]. IoT devices can improve various aspects of daily living, health, and well-being. However, some devices have limited security features which potentially expose consumers to cybercrime threats. One of the reasons that such devices may have weak security is due to their lightweight nature. That is, during the development lifecycle of many IoT devices, the focus is on the functionality of the device rather than the security of it. As a result, attackers can exploit vulnerabilities to include those present in IoT device sensors or communications protocols, or they may take advantage of weak or absent encryption. Reviewing the security of IoT devices is essential as they become increasingly embedded in daily life and people's routine activities [2], which creates new opportunities for criminals to exploit.

According to a survey conducted in 2019 [3], 65% of consumers believe that hackers monitor their IoT devices and that 60% of their data is leaked when these attacks are carried out which facilitates many dangerous attack vectors. Previous work [4] [5] has reviewed the threats posed by consumer IoT devices but, as discussed, the threat landscape is evolving and hence it is necessary for researchers to keep on top of this. Consequently, this paper presents a systematic review that incorporates the most recent literature regarding security attacks and crimes that have the potential to be facilitated by connected devices.

The paper is organised as follows. Section 2 provides the background to the paper, covering the protocols and processes attackers exploit when targeting IoT devices. Section 3 explains the methodology of our systematic literature review and the steps taken for the analysis. Section 4 presents our findings, while Section 5 provides a discussion of the results. The last section focuses on new avenues for research in this field.

## 2. BACKGROUND

This section provides an overview of the protocols and processes used in IoT environments and that are of particular interest to attackers targeting these devices.

### 2.1 How data communications are modelled over the Internet for consumer IoT devices

At present, two conceptual frameworks exist for modelling communications over internet-connected devices. These are the Transmission Control Protocol (TCP) / Internet Protocol (IP) Stack and the Open Systems Interconnection Model (OSI) Model. The TCP/IP Stack consists of five layers, while the OSI Model consists of seven. As the OSI Model is more detailed, we use this model here to discuss the potential points at which attackers might target IoT devices. The OSI model extends the TCP/IP stack by describing the process of data communication all the way from the physical layer (the physical device's Network Infrastructure Controller) to the application layer (where the services accessed by the device exist, such as email, internet, gaming, voice and video calls) The seven layers of the OSI model are shown in Table 1. As will hopefully become apparent, the OSI model and its associated layers provide a useful framework for discussing the variety of security attacks (which may target different OSI layers) that may be used to target IoT devices and for organising the results of the systematic review. In terms of the OSI communication process itself, this starts at layer 7 (the application layer) and moves through the other layers, ending at layer 1 (the physical layer where data is modulated through a device such as a modem to its intended destination). To take an example, consider a scenario where a device wants to access a website (to download information from it). Table 1 shows the steps taken at each layer.

| OSI Layer | Description |
|---|---|
| 7 – Application | The layer where the specific application and/or service is provided and accessed by the user. In the context of a smart doorbell this would be where (say) a video conferencing application or service resides. Example protocols that facilitate video and audio communication for doorbells include the Real Time Streaming Protocol (RTSP), Service Initiation Protocol (SIP), Hypertext Transfer Protocol (HTTP), Hypertext Transfer Protocol Secure (HTTPS), Voice-over-IP (VoIP) Protocol and Domain Name System (DNS). |
| 6 – Presentation | The layer where the information is secured appropriately for the specific application/service to be accessed at the Application layer. This is typically where encryption occurs (however, in the context of HTTPS, this also occurs at the Application layer). In the context of a smart doorbell, data packets communicated between the recipients should be end-to-end encrypted using protocols such as the Secure Socket Layer (SSL), Transport Layer Security (TLS) and the Hypertext Transfer Protocol Secure (HTTPS). |
| 5 – Session | The layer where a socket connection is established ready for data packets to be encrypted at the presentation layer. The sockets that facilitate connectivity between recipients consist of both SOCKS4 and SOCKS5 protocols. |
| 4 – Transport | The layer where information is prepared for transport up the OSI stack. This is achieved using two main types of protocols. These are the Transmission Control Protocol (TCP) and User Datagram Protocol (UDP). In the context of a smart doorbell, video communications are often transmitted via UDP. UDP is considered a connectionless protocol – meaning a connection does not need to be established before sending the information. This decreases transmission time due to the reduced communication overhead but makes the connections less reliable due to the potential for packet loss. Examples of protocols that run over UDP for smart doorbell communications are the Service Initiation Protocol (SIP) and Voice-over-IP (VoIP). For TCP, examples include the Real Time Streaming Protocol (RTSP), Hypertext Transfer Protocol (HTTP) and the Hypertext Transfer Protocol Secure (HTTPS). |
| 3 – Network | At this layer, information is gathered from Layer 2 (where information is only transmitted between devices present on the local network – a process known as Switching) where the packets are then Routed using Layer 3 across networks using a **Router** (where information can then leave the network – a process known as Routing). The Internet Protocol (IP) also operates at this layer. As packets are Routed across to another network, the packets are passed to the Transport Layer where either the TCP or UDP packet headers are appended to the frame. In the context of a doorbell, this would encompass a protocol such as VoIP or SIP where the UDP header would consequently be appended to the packet. |
| 2 – Data Link | This layer manages the transmission of data for each step along the route between the source and destination by switching packets of information using a **Network Switch** or the Address Resolution Protocol (ARP) so that packets can be passed to the Network layer to be routed ready for transmission up the OSI stack. |
| 1 - Physical | At this layer information is transmitted from the device's physical network communication system (i.e. network interface card, Wi-Fi, etc.) as a signal that arrives at the next hop in the routing process (i.e. over a WAN Link to the Internet via a device such as a **Modem** or from a device via Wi-Fi to the Wireless Access Point or ethernet). In the context of a doorbell this would be how the Doorbell would physically communicate with the local network, usually via Wi-Fi or ethernet. |

*Table 1 OSI Model Layer Descriptions.*

According to [6] people perceive that cybercrimes towards IoT devices already exist. According to Marton 2023 [7], in the first six months of 2023, IoT malware was up by 37% resulting in a total of 77.9 million attacks, compared to 57 million attacks in the first six months of 2022. As such, this review will focus on the following research questions:

RQ1: What cyber-attack vectors are possible using IoT devices?
RQ2: Which consumer IoT device platforms can or do these criminal activities take place on (e.g., smart watch, TV, smartphone, etc.)?
RQ3: Can attacks be mitigated or their likelihood reduced? If so, how and what is the prevalence of studies in the information security community?
RQ4: Which attacks against IoT devices can be used to commit crimes?

## 2. METHOD

Unlike ad-hoc literature reviews, systematic reviews follow a structured methodology, the aim of which is to produce a focused and unbiased synthesis of the relevant literature. In what follows, we follow PRISMA guidance [8] and specify the search terms used, the electronic databases searched, the inclusion and exclusion criteria applied, the data extracted, and the method of synthesis.

### 2.1 Electronic Searches

The following electronic databases were searched in July 2025: ACM Digital Library, Directory of Open Access Journals (DOAJ), IEEE Xplore Digital Library, ProQuest, Scopus, and Web of Science. Searches were limited to papers published between 2012 and 2025. This timeframe was used to confine the review to contemporary issues. The search terms were piloted to achieve an acceptable balance between finding relevant and irrelevant articles. To identify the appropriate search terms to conduct the electronic database search an initial Google search was conducted to identify different acronyms and descriptions of consumer IoT devices. This was informed by the search terms used in a previous review [4] but with some modifications. This gave an overall first impression of the keywords and potential security attacks to include in the search strategy, (e.g. exploit, vulnerability, hacking, attacks, malware, IoT, Fog, Edge). The search terms were then piloted in an iterative fashion to achieve a balance between sensitivity (retrieving a high proportion of relevant articles) and specificity (retrieving a low proportion of relevant articles). The final search terms used were as follows:

("Internet of Things" OR "smart wearable" OR "smart device")

AND

("hack*" OR "threat*" OR "software vulner*" OR "attack*" OR "crim*" OR "exploit"
If the databases allowed the following search terms were used instead of the associated search terms above: SU.EXACT("cyber hack") OR SU.EXACT("cyber threat") OR SU.EXACT("cyber attack") OR SU.EXACT("cybercrime"))

AND

("consumer" OR "smart home"
If the databases allowed the following search terms were used to ensure that only consumer IoT smart devices were shown and no industrial or healthcare-based IoT devices are shown: SU.EXACT("consumer") NOT "medical" NOT "industr*" NOT "health" NOT "business" NOT "commercial" NOT "healthcare" NOT "computer" NOT "patient")

The SU.EXACT syntax ensures that only exact matches to specified search terms are identified. Not all search engines allow the use of this syntax, and so various forward and backward searchers were used. Consequently, the search terms were modified, where necessary.

## 2.2 Inclusion & Exclusion Criteria

A fundamental part of the systematic review process is to define the inclusion and exclusion criteria used to select articles for analysis. This was completed by using a variation of the PICOS criteria [9] commonly used in systematic reviews. PICOS stands for Population/problem, Intervention(s), Comparator, Outcomes and Study design. Even though Crimes are considered a "Problem" in the original PICOS criteria, it was still the belief that for the purposes of this review that this would not be the appropriate term to be used, as this review largely focuses on vulnerabilities to consumer IoT devices. As such, this is modified to DICOS (See Table 1) with the population criteria being substituted with "Devices" to better suit the purposes of this systematic review.

| Criteria | Inclusion | Exclusion |
| --- | --- | --- |
| Devices | Consumer Internet-of-Things (IoT), Internet-of-Vehicles (IoV), Personal Internet-of-Drones (IoD), routers, games consoles, smart watches, Internet-of-Medical-Things (IoMT) devices (but limited to wearable health monitors, smart watches, glucose monitors, etc.) and smart TVs. | IoMT (specifically Pacemakers, Ultrasound Scanners, X-Ray Machines), Critical Infrastructure IoT devices, Smart Cities, Laptop and Desktop Computers. |
| Intervention(s) | Not applicable. | Not applicable. |
| Comparator | Not applicable | Not applicable |
| Outcomes | Security attacks, device vulnerabilities, malware, crime types and sub-types. | Papers discussing security attacks that are limited to the English language only. |
| Study design | Peer-reviewed journals, conference proceedings, survey papers, systematic reviews, government documents or academic thesis only.<br><br>Any paper that includes qualitative and quantitative data collection or a mixture of them. | Papers that have not been subjected to the peer-reviewed process. Papers that are behind a paywall that the authors institution does not have access to. These documents were not included as well: Commentaries, Forewords, Books/book reviews, Articles, Opinions, Letters, Editorials |
| Other | English language. | Non-English. |

*Table 2 A summary of the eligibility criteria for the screening phases.*

Identified citations were imported into the reference management software *EndNote* and duplicates removed. Articles were then screened using the PICOS criteria on the basis of their titles and abstracts. To ensure replicability and that the inclusion/exclusion criteria were being correctly applied, an Inter-Rater Reliability (IRR) exercise was conducted. To do this, the titles and abstracts of 10% of a random selection of studies were screened by two coders. IRR was calculated using the Cohen's Kappa Formula [10], and the resulting value of 0.70 indicated an overall acceptable level of agreement between the two coders. The full text of those that met the inclusion criteria were then read and the articles assessed by a single coder.

## 2.3 Data extraction and management

A pro forma, piloted on a sample of articles to ensure that relevant information was captured [11], was developed to extract the following information from each study:

- Year of study
- Publication type
- Study design
- Quality of evidence
- Type of evidence (e.g., empirical or simulation)
- Target of crime, method of offending, cybercrimes/harms
- Type of device
- Brief description of study

Studies vary in terms of the research methods that they employ, with some employing more rigorous approaches than others. For this reason, an assessment of methodological rigour is a common feature of systematic reviews, and various hierarchies of evidence have been used to do this. Table 2 shows the hierarchy of evidence, or hierarchy of feasibility used in a previous review [4] and the one that is adopted here. To take an example, a security attack to a consumer IoT device is considered more feasible if that attack has been demonstrated in the real world than an attack that researchers have merely speculated about.

| Hierarchy of Evidence | Type of Evidence |
| --- | --- |
| Real world | Paper demonstrates a specific attack vector, vulnerability (e.g. an identified security weakness presents in a device's firmware), exploit (where an attacker or security researcher has exploited a vulnerability in a device) or malware attack against a consumer IoT device and the resultant consequences on real consumer IoT systems. |
| Experimental (lab-based) | Paper demonstrates a specific attack vector, vulnerability, exploit or malware against a consumer IoT device and the resultant consequences in a lab-based technical experiment but said attack is confined strictly to the lab-based experimental situation. |
| Experimental (simulation) | Paper demonstrates a specific attack vector, vulnerability, exploit or malware against a consumer IoT device in a computer-generated simulated exercise. Examples include the testing of datasets against a proposed mitigation strategy with the results presented in a simulation graph. |
| Expert speculative | Attacks, vulnerabilities, exploits or malware, and resultant crimes are speculatively derived by a group of technical experts and researchers. |
| Author speculative | Attacks, vulnerabilities, exploits, malware, and resultant crimes are speculatively derived by the author of a particular study. |
| User speculative | Attacks, vulnerabilities, exploits, malware, and resultant crimes are speculatively derived by a group of users of consumer IoT devices. |

*Table 3 Hierarchy of evidence/feasibility used to assess studies (see [4]).*

Having identified the papers, a thematic analysis was employed to synthesise the findings. Thematic analysis is an inductive or deductive approach for extracting recurrent themes in text data, which is commonly used in qualitative data analysis [12].

## 3. RESULTS

### 4.1 Summary of search results

Figure 1 shows that the initial searches identified 12,705 articles. After the removal of duplicates and following the screening of titles and abstracts, 9,064 articles remained (i.e., 8,032 were excluded). Excluded studies either discussed consumer IoT but not security or discussed security but not consumer IoT security attacks. An additional 488 studies were excluded during full text screening, leaving 544 studies for synthesis. All included studies were coded by the first author of the paper.

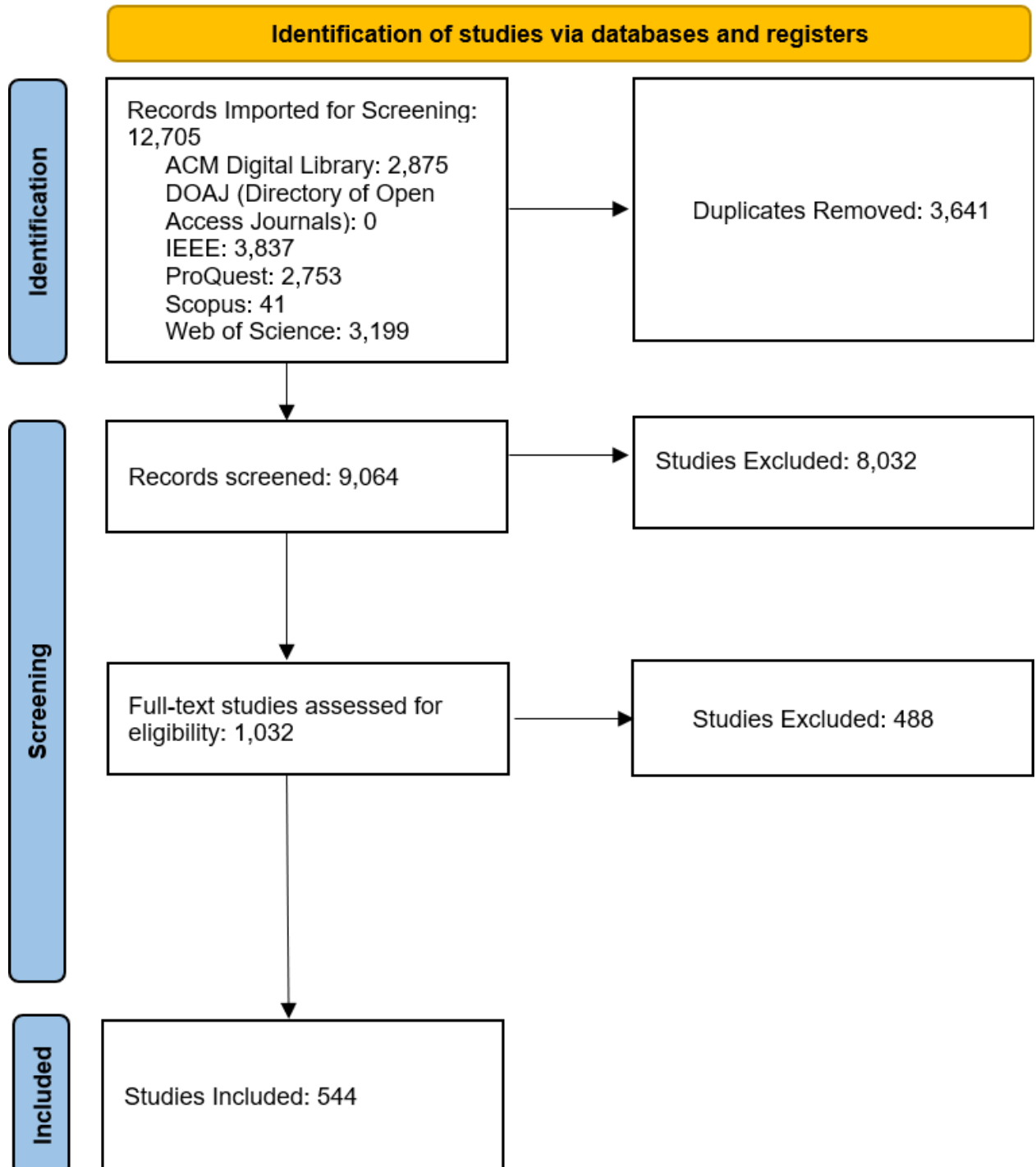

*Figure 1 PRISMA Diagram.*

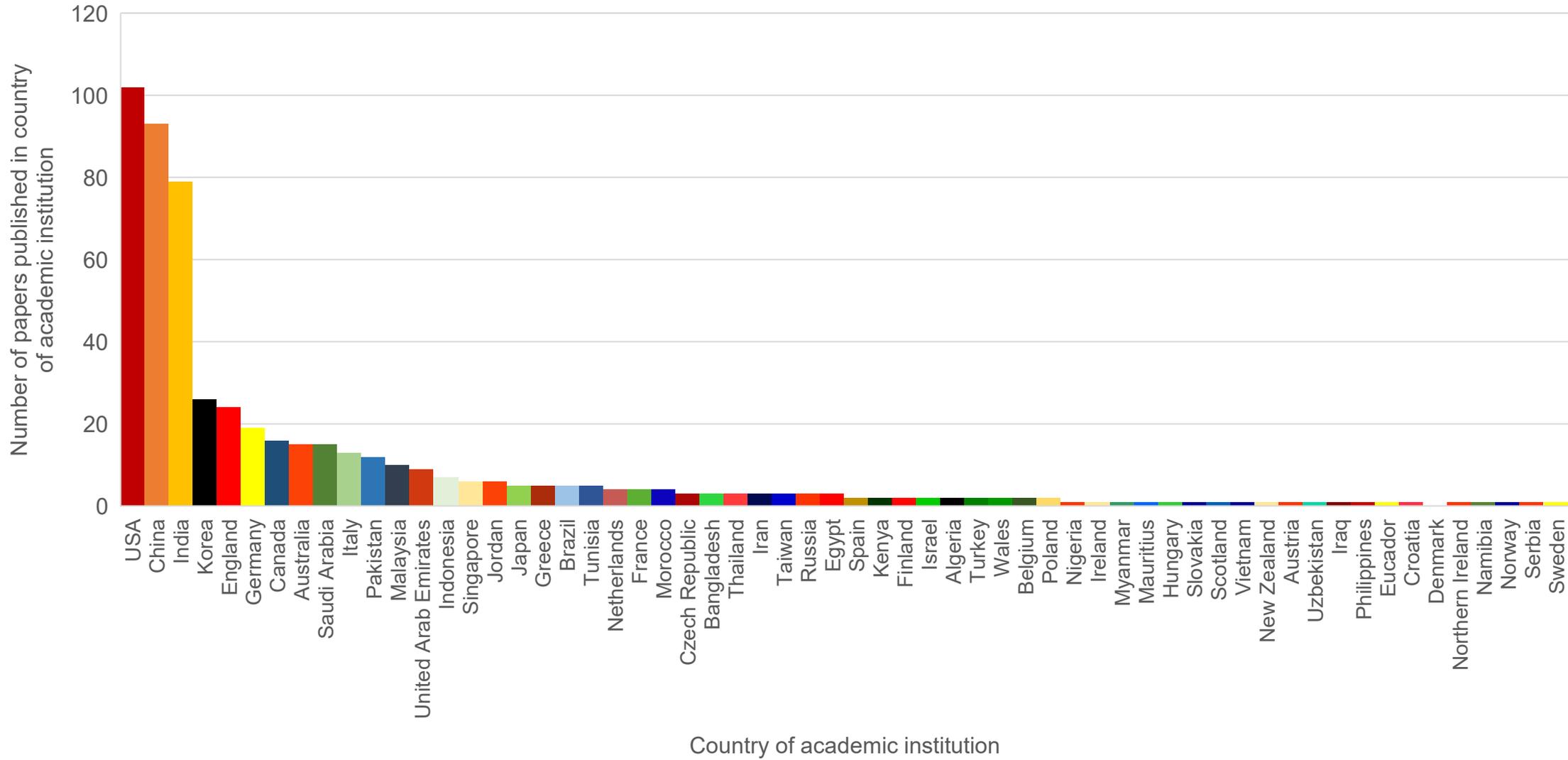

*Figure 2 Author countries of published papers from results.*

Before analysing the findings, we examined the geographical locations of the academic institutions that the authors of the papers were associated with. Where a study involved multiple authors from different academic institutions, the country associated with the lead author's academic institution was used. Figure 2 shows that the included studies were written by authors from a wide range of countries. However, papers were more likely to be written by authors from the USA (102 out of 543 papers), China (93 out of 543) or India (79 out of 543).

**4.2 Attacks against consumer IoT devices**

The literature reviewed shows a wide range of attacks and behaviours. In some cases, the attacks were adapted to the type of device and in others they depended on the threat actor's goals. According to [13], the most common attacks to consumer IoT devices can be grouped into four categories – those that target **devices**, **networks**, the **cloud**, or **applications**. We use these categories here. Figure 3 shows a Sankey diagram that summarises the cyber-attacks discussed. Of the 543 selected papers, 253 were focused on attacks, using different methodologies and focusing on different aspects of the operations of these devices. Figure 3 shows that the vast majority (74.31%) of papers focused on network attacks. Most (N=188) papers that did so focused on vectors that were relevant to any IoT device, while the remaining 65 papers focused on issues and vulnerabilities that were specific to devices or groups of them.

Considering the OSI layers to which the attacks apply, layers 2, 6, 5 and 4, had similar numbers of papers focused on them. While layers 7 (N=74), 1 (N=55) and 3 (N=32) were the top three that scored the highest numbers of studies, respectively. While Layers 6 (N=19), 2 (N=13), 4 (N=9) and 5 (N=4) scored the lowest. In terms of the methodologies used, it appears that most papers employed 112 Experimental (Lab-based) environments, 71 Expert Speculative environments and 51 Experimental (Simulated) environments. Author Speculative (N=13), Real-world (N=5) and User Speculative (N=1) environments all scored the least, respectively.

We have created a table to show the detailed division of the 253 papers into the hierarchy of evidence classification. This table is the full coding of studies from the resultant papers of the systematic review. Due to the size of it, it is available as extra documentation online[1] named Table H1. The level of detail of such table does not provide further information to the scope of the paper but allows full reproducibility and can be of use to the readers to further explore specific areas for their own research purposes.

---

[1] Supplementary Material can be accessed at:
https://drive.google.com/drive/folders/1UmYRciO49Bmw7PzmvX90XkX-K6_UHydn?usp=sharing

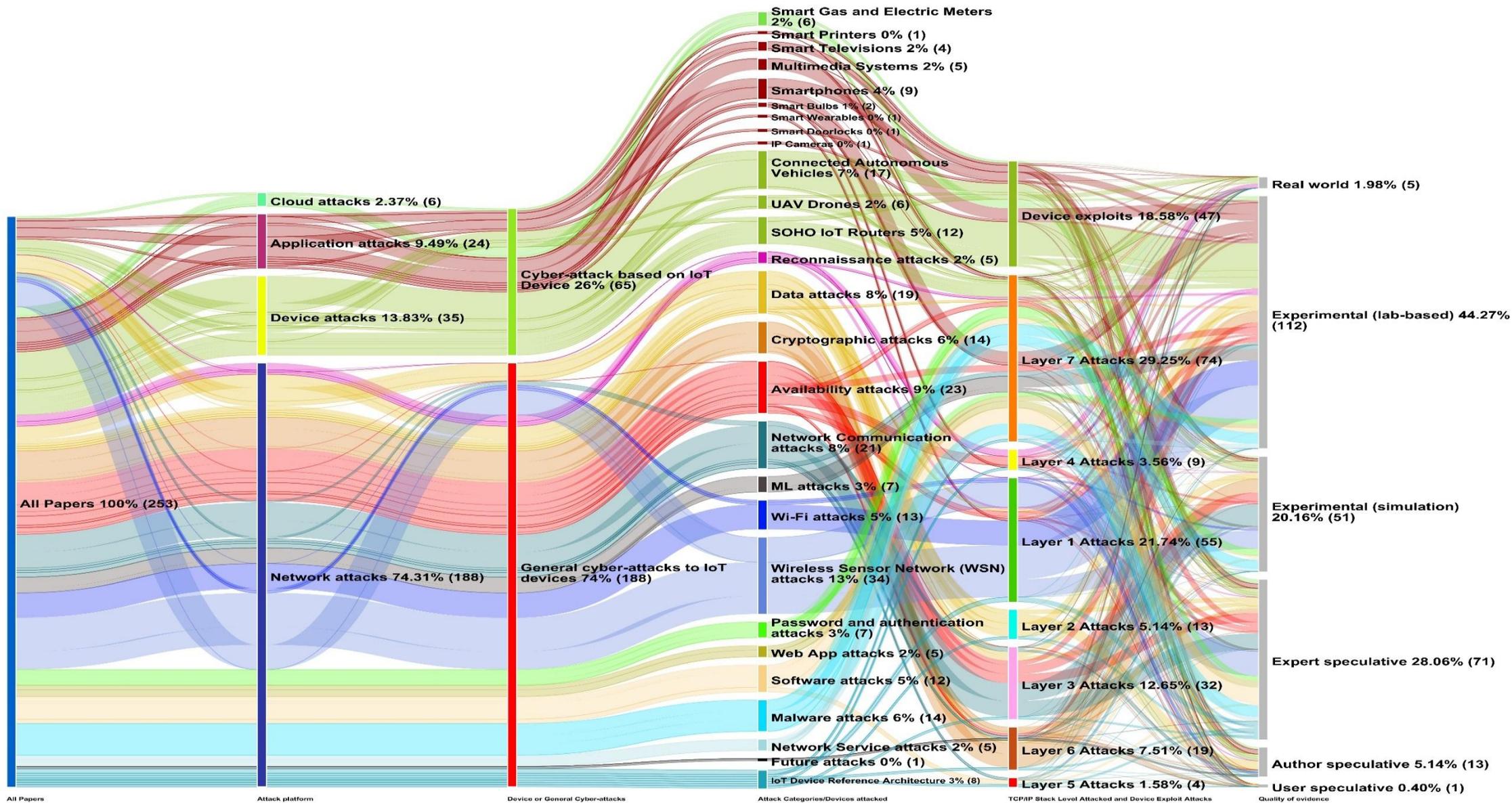

*Figure 3 Sankey Diagram to illustrate studies for security attacks.*

### 4.2.1 Attack categories

The remaining works were divided among Device attacks (29), Application attacks (16), and Cloud attacks (6).

As discussed, network attacks were the most common type identified (132 out of 186). Network attacks are those that happen without infecting the device, but using vulnerabilities related to the implementation of the OSI layer protocols. This is a rather large umbrella that contains attacks against the Physical link [14] [15] [16] [17] as well as the IP protocol [18] [19] or even the application layer of the OSI stack [20] [21]. Some of these attacks involved the presence inside the device network through malware infection or the physical presence of the attacker [22] while others could be undertaken remotely [23]. There were too many articles to include citations, or to provide further details about them here.

Device attacks – which were discussed in 29 papers – describe attacks that target the devices themselves in different ways. Some of the attacks identified exploit flaws in the device design through applications present on the device [24]; in other cases attackers can exploit firmware– [25] or hardware–vulnerabilities [26]. These types of attacks can allow a full takeover of the device by cybercriminals, enabling an attacker to commit various types of crimes. Figure 3 shows that the targeted devices discussed were limited to vehicles, drones, and routers.

Application attacks (N=16) are those that target the applications of on certain IoT devices without attacking the devices firmware or hardware vulnerabilities. These types of attacks, much like device attacks, can allow an attacker to take over the device to facilitate criminal activities. Figure 3 shows that the targeted devices discussed in the selected literature are smart TVs [27], smart multimedia systems [28], smartphones [29], smart printers and smart doorbells [30].

Cloud attacks, which were discussed in only six papers, target systems that require the use of the cloud for an IoT application to operate. In the literature, these were identified as attacks that targeted smart gas and electric meters [31].

### 4.2.2 Attacks categorised through the OSI stack

To properly categorize the different general consumer IoT device attacks, attacks were categorized against the OSI stack. Once attacks have been successfully categorized against each OSI layer, each subsequent subsection for each OSI layer will identify how authors in the literature conducted experiments or surveyed these attacks, alongside the Hierarchy of Evidence as a point of reference to help the reader understand how researchers tackled these research activities.

Figure 3 shows that while Layers 4, 5, and 6 of the OSI stack were examined by researchers, this was only regarding DDoS attacks (for Layers 4 and 5), and Network Service attacks (i.e. DNS attacks for layer 6). Layer 2 was also of less interest to researchers than the other layers. This may be due to the relative gains in exploiting these layers with respect to the effort involved; for instance, devices may well be vulnerable to DDoS attacks like SYN floods [32] or UDP [33] (layer 5) or using ICMP [34] (layer 3) but targeting a single device has reduced impact that may not justify the cost to cybercriminals.

**Layer 7** was the most studied layer (53 studies). This layer manages applications that consumer IoT devices use for specific functionalities or applications. It is perhaps for this reason that studies that examine this layer focus primarily on Malware attacks such as Botnets [21], Trojans [35] or Ransomware [36] as these different types of malware target device applications (i.e. firmware or software) through infection.

| Attack | Description |
| --- | --- |
| HTTP Flooding attacks | The process of an attacker flooding the network and/or device with partial Hyper-Text Transfer Protocol (HTTP) website GET Requests to trick the device into believing that the requests are legitimate website User-Agents when in fact they are malicious attack attempts. This then allows the attacker to circumvent a Firewall, Endpoint Security, or Intrusion Prevention System (IPS) as the device or network will believe that these packets are legitimate website access attempts to result in a Denial-of-Service (DoS) attack. |
| Flooding attacks | An attack whereby an adversary uses various protocols such as TCP, UDP, HTTP or ICMP to overwhelm the device and/or network with packets of these protocols to the point that it affects network traffic and/or exhausts device and/or network resources such as CPU usage, Memory, or Network Bandwidth to the point that normal network operations are adversely affected or cannot be used at all. This effectively facilitates a DoS situation as it affects device and/or network availability. |
| Reflective Distributed-Denial-of-Service (DDoS) attacks | Also known as Amplification DDoS attacks. These attacks are facilitated by adversaries either exploiting publicly used Internet Application Services such as Domain Name System (DNS) or Network Time Protocol (NTP) public servers. Amplification DDoS attacks occur either through DNS Amplification attacks or through the NTP Service. By spoofing the source IP address of these services, attackers can amplify traffic and disguise their identity at the same time. |
| Data poisoning attacks | The process of an adversary injecting fake data into the training set of machine learning systems. Data Poisoning is the most common form of Adversarial Machine Learning. The attack directly targets the IoT devices data analytics to ensure that data collected from the sensors is either improperly processed by the data decision layer or the decision layer takes the incorrect action based on the poisoned training sets due to manipulated training data. |
| Evasion attacks | Malicious samples are sent directly to the training sets at training set test time within the Machine Learning model present on the IoT device to result in the incorrect results being output from the device attacked. |
| Worms | A type of malware that deliberately injects itself into the code of the system via its communication interface. Once there it begins to replicate itself endlessly and then attempts to infect other devices on the Local Network, where the process on the originally infected device begins again. |
| Trojans | Trojans are defined as a type of malware that are disguised as legitimate software applications but that are malicious. It was shown that Using Routers and Network-On-Chip (NoC) Devices, of which routers are the most common to feature NoC, it is possible to launch a powerful DDoS attack from a Trojan Malware that infects NoC Devices. Obfuscation is a powerful tool that is used by adversaries to make the Trojan look legitimate. |
| Backdoor | A type of malware that is obfuscated using a Trojan. In this attack the adversary delivers a malicious file, sometimes called a Payload onto the victim device via social engineering methods. The victim opens the file, believing the file to be legitimate due to its obfuscation and then the malware creates a tunnel to the attacker from the victim device. This then allows an attacker a direct entry point into the device, circumventing and side stepping all security mitigations such as Network Files and Endpoint security, to execute a host of malicious actions. |
| Botnets | Botnets are described as a type of malware that turns endpoints/nodes into slave or Zombie Machines communicating with a Malicious Command and Control Server (C & C). |
| Ransomware | Ransomware Assaults are usually actuated through a Trojan. It usually enters through an ignorant click of the victim from a link attached in an email or unsecured exposure in the network. The main goal is to infect the victim's system and encrypt all the victim's files on the device that's become infected and potentially other devices connected to the same network. Once all the files have been encrypted the Ransomware demands a ransom in return for the successful restoration of the victim's files. |

| DNS Poisoning attacks | Also known as Domain Name System (DNS) rebinding attacks. This attack allows a threat actor to redirect communications to a malicious entity rather than data going to its correct destination. The attack works by exploiting either the DNS A records or time-varying DNS. DNS is responsible for translating website URL addresses to remote Layer 3 Network IP addresses. |
|---|---|
| DNS Water Torture attacks | The process of disrupting the Domain Name resolution at the server-side of the Server that contains the DNS A Name Resolutions that the victim IoT device is connected to. This enables the attacker to prevent the Domain Names that IoT devices are translating to at the Client-Side from reaching the resolved IP addresses at the Server-Side. |
| DNS Amplification attacks | The attacker spoofs the IP address of the victim to cause a Denial-of-Service (DoS) attack. The attacker sends a small request to the DNS server and then the DNS Server responds with a large reply. To achieve a high impact the query type ANY is used by the attacker to return all information of the victim. |
| IoT Device Fingerprinting attacks | Attackers aggregate network traffic data sets through network telescopes, honeypots, and similar collection architectures. This allows attackers to identify IoT device specific information such as Make, Model, Brand, Communication technology, Software/Firmware version, Identified Vulnerabilities, Malware vulnerabilities, etc. |
| Brute-force attacks | A trial-and-error method of obtaining a password or a key to an encryption algorithm. The attacker tries every single possible combination until the appropriate password or key is found. This is a very time-consuming approach to obtaining user credentials, however, once the password or key is identified the attacker is then able to gain unauthorized access to a communication stream between a device and the destination, or unauthorized administration access to an account on a particular device. |
| Dictionary attacks | Most often used for offline attacks but can be used for online attacks as well. The attacker uses a pre-determined list known as a dictionary wordlist to try each password until the correct password is identified. With the knowledge of (l-1 password shares) the attacker can try each password present in the share until the password is identified. The advantage of using this method to crack a password over the Brute-force method, is that because you are using a pre-determined wordlist. If the password is present in the list, then the password can be found that much quicker. Many attackers try this method of password attack first before moving onto the more radical Brute-force attack that takes significantly longer, as it must try every single password combination. |
| Hard-coded Credential attacks | Hard-coded credentials are administrator-level credentials that have been added to a consumer IoT device by the manufacturer without the knowledge of the victim. As such this presents a significant vulnerability for the attacker to exploit as it allows administrator-level access to an IoT device without the consent or knowledge of the victim. Hard-coded credentials, weak or guessable passwords are in the Open Web Application Security Project's (OWASP) Top 10 security flaws in IoT devices. |
| Known-key attacks | The attacker eavesdrops a wireless communication medium and extracts previous session keys from the IoT device. If these old session keys do not use timestamping to make each key no longer valid, the attacker can use these session keys to execute a known-key attack, whereby the attacker uses an old key to gain unauthorized access to the IoT device. |
| Privileged-insider attacks | The malicious entity uses the registration information of the legitimate users during the session registration phase of the session authentication, of which is sent to the Registration Authority (RA) to gain unauthorized insider, administrator and even root user level privileges to a device and consequently additional devices inside the consumer Smart Home Local Area Network (LAN). This attack is also known as a Privilege-Escalation attack and allows the attacker to exfiltrate data, change file permissions, exploit vulnerabilities and act as a vector/gateway to further additional attacks on the Smart Home network. |
| Social Engineering attacks | In these attacks victims are humans instead of networked devices. Users are attacked psychologically. The adversary attempts to communicate directly with the user, through pre-existing attack methods such as Phishing or Malware execution, and attempts to provoke the user by tricking them into believing what they are clicking on is genuine and emergent. |
| SQL Injection attacks | Attackers target the web application database of a client or server through a preexisting vulnerability. Malicious characters are added to an SQL query by the attacker that ensures that certain strings of characters in the database query always convert to true, even when by normal standards, it should convert to false. |
| Cross-site Scripting (XSS) / Cross-site Request Forgery (CSRF) attacks | XSS/CSRF attacks are facilitated by an attacker writing scripts in web applications that will lead to attacks targeting specific web applications. The main goal is to use special characters to make the browser interpreter switch from a data context to code execution that will perform a misuse in the victim's application. While XSS primarily attack Web Browser Applications in desktop workstations, smartphones, they can be present as vulnerabilities on specific devices firmware/software as well as any associated applications. |
| Malicious Code Injection attacks | Also known as a False Data Injection attack (FDIA). In this attack the adversary uses a malicious node present on the network, or in some instances off the network, to inject malicious data such as a faulty state into the devices firmware, software or other runtime applications of a smart device such as smart meters, smartphones, tablets, etc. to either exploit a vulnerability, manipulate a devices functionality or overwhelm a device to the point that it no longer functions at all or functions in a malicious manner, for example, if an attacker where to inject malicious data into a smart meter, the attacker could manipulate energy tariffs affecting consumer energy costs. |
| Buffer Overflow attacks | The attacker uses Assembly Code to overflow the Stack (Buffer) within the memory (RAM) by ensuring that the return address points to arbitrary code that has been injected by the Buffer Overflow attack to ensure that the stack overflows within the memory to allow for the execution of vulnerable exploits or to perform a Denial-of-Service (DoS) attack. |
| Fuzzing attacks | A very old technique that is used by software developers and IoT hardware device diagnostics companies to detect Hardware issues for the purposes of diagnostics or for testing Software programs. While Fuzzing is mainly used for non-malicious purposes by authorized individuals to do so, in recent times Fuzzing attacks have emerged that allows an adversary to Reverse Engineer the system to facilitate a *Reverse Engineering attack*, for the purposes of exploiting hardware or software vulnerabilities. |
| Reverse Engineering attacks | The process of an attacker using Fuzzing-based methods to survey an IoT device for Physical Hardware and/or Critical Software Vulnerabilities to exploit through another attack. The most common way for an attacker to facilitate this attack is through *Fuzzing attacks*. |
| Booting attacks | In edge devices, built-in security mechanisms do not work at the time of the boot process. During this process, devices become more vulnerable to various security attacks. Attackers take advantage of this weakness and target devices for their malicious purposes. A booting attack is applied at the start of the system when devices are getting ready to communicate or security algorithms are not installed yet. Common booting attack protocols used are UART or JTAG. |
| Data exfiltration attacks | Data Exfiltration is defined as the situation whereby data, often very sensitive data, is leaked out of a network to an attacker than can then use this data for malicious purposes. |
| Data Forgery attacks | The process of a threat actor that involves the manipulation and tampering of data as it is communicated from the device to router and across the external network. Forged or tampered data can cause major issues with IoT device functionality. For example, if a software update was tampered with, the code in the software could be rewritten by an attacker to perform additional malicious functions such as slowing the device down, exfiltrating data or various other potential illicit actions. |
| Message Queueing Telemetry Transport (MQTT) attacks | A lightweight unencrypted protocol that enables lightweight network communications between lightweight consumer IoT devices. Attacks that can be used to exploit the unencrypted MQTT protocol include Man-in-the-middle (MiTM), Eavesdropping and Data Exfiltration attacks. |
| E-mail Spam attacks | The attacker sends an unwanted e-mail or message to the consumer IoT device potentially harbouring unwanted software such as Malware to obtain personal information about the victim and/or device and degrade the service of that device. |
| E-mail Phishing attacks | The attacker pretends to be a trusted or legitimate entity to trick the victim users and convince them to provide their sensitive data or click on a malicious link |
| Web Directory Brute-Force attacks | The attacker brute-forces directories that may be present from an advertised Hypertext Transfer Protocol (HTTP)/ Hypertext Transfer Protocol Secure (HTTPS) web service(s) on any device including consumer IoT devices |
| DNS Flooding attacks | The attacker floods the DNS server with multiple malicious DNS queries with the intention of facilitating a DDoS type situation by preventing the DNS server from sourcing DNS IP address to Web Address URL Fully Qualified Domain Name (FQDN) translations. |
| Extensible Markup Language (XML) Signature Tampering attacks | An attacker targets the XML signature files that IoT devices use to authenticate the data and integrity of the information received. This then subsequently, allows the adversary to tamper with any information sent from an IoT device. |
| Session Initiation Protocol (SIP) | The adversary facilitates a Denial-of-Service (DoS) attack when the attacker floods the advertised SIP protocol by sending numerous SIP INVITE or REGISTER packets to overwhelm the SIP service to prevent normal video conferencing taking place through the SIP protocol. |
| Flooding attacks | |
| Rootkits | An attacker installs harmful software on the systems devices to take control of the applications they oversee. |

*Table 4 Layer 7 - Application Layer attacks*

**Layer 1** was the second most studied layer (35 studies). These attacks focus on the physical layer of a device, which can consist of sensor technologies [37], device software [38], Wi-Fi technologies for initial communications [39], Password attacks [40] and IoT device web applications [41]. The variety of technologies that can be attacked, and the opportunities that they provide, may well explain why this layer was as popular as it was for research in this area.

| Attacks | Description |
|---|---|
| Jamming attacks | The aim of a jamming attack is to disrupt the physical layer wireless communication frequencies between the smart device and the hub. To achieve this the attacker exploits a high-power radio source to emit wireless signals with the same working frequency as the signal that is being emitted by the associated physical layer protocol from the device and recipient, resulting in a Denial-of-Service (DoS) attack. |
| Wi-Fi Eavesdropping attacks | The process of an attacker passively eavesdropping leaked side-channel data of Wi-Fi networks. Wi-Fi networks can leak information through side-channels facilitating data privacy leakage. From these side-channels attackers can sniff packets to obtain sensitive information, without even needing to be present on the victims Wireless (Wi-Fi) network. |
| Wi-Fi De-authentication attacks | Adversary facilitates a wireless evil twin attack by first forcing the consumer device off the home network. This is done by using malicious information on the wireless medium to force the victim device to disconnect from the legitimate wireless network |
| Wi-Fi Re-authentication attacks | Once the adversary has successfully de-authenticated a victim's device off the legitimate router, they then force the victim device onto a wireless network controlled by the adversary. |
| Wi-Fi Evil Twin attacks | The attacker de-authenticates and then re-authenticates the victim onto a malicious wireless network controlled by the adversary. This will have the same Network Name; sometimes also referred to as the Service Set Identifier (SSID), and various other wireless network properties to maintain persistence and act as a gateway to conduct any of the attacks listed in this table with ease due to the victim device being present on the malicious network. |
| Wi-Fi Jamming attacks | The process of an attacker attempting to use a Wi-Fi device (such as the Alfa AWUS036h) to jam the signal on a specific Wi-Fi Channel to create attenuation, noise, and delay on that channel to either reduce the availability of that channel between the IoT device and the Wi-Fi Access Point or to restrict it entirely and as such facilitating a *Jamming attack* (See Layer 3, Availability attacks). As such this attack can completely disable the devices communications between the device itself and the internet and as such facilitates a Denial-of-Service (DoS) attack. |
| RFID attacks | Radio Frequency Identifier (RFID) tags are used to identify everyday objects, which enables the tracking ability of objects throughout space and time in a sustainable manner. They have a wide range of applications such as electric toll collection systems, access management systems, airport baggage tracking logistics and vehicle Remote Key Entry (RKE) systems. Attacks to RFID include the *Full Disclosure attack*, *Tag Removal and Destruction*, *Temporary Paralyzing attacks*, *Kill Command attacks*, *Sybil attacks*, *attacks on RFID readers*, *Unauthorized Tag Reading*, *Tag Modification attacks* and *Software attacks*. |
| ZigBee attacks | Examples of ZigBee attacks include the ability to compromise other devices on the network facilitated by a *ZigBee Concealed Wireless Jamming attack*, the delivery of malicious payloads or commands to devices using the ZigBee protocol facilitated by a *ZigBee Passive Inference attack* and the eavesdropping of data between the ZigBee device and the ZigBee Hub through a *ZigBee Waveform Emulation attack*. |
| 802.15.4 KillerBee attacks | A framework of tools that allow an adversary to exploit vulnerabilities in 802.15.4 devices such as ZigBee, LoWPAN and Thread. KillerBee simplifies sniffing, injecting traffic, packet decoding and manipulation, as well as reconnaissance and exploitation. Numerous other attacks can be carried out using KillerBee such as PANId conflict, replay attacks, packet capturing and network key sniffing. |
| NFC Replay attacks | This attack is conducted with the use of a device known as a Mole. A Mole is an adversarial machine that allows an attacker to potentially intercept the payment information from a smartphone while the victim is using the NFC system on their smartphone to pay for an item to Replay the Payment Information to their Mole so that they can capture the payment details and potentially use this information to pay for items later. |

| Z-wave Side-channel Analysis (SCA) attacks | The process of an attacker using a wireless packet sniffer such as the Software Defined Radio (SDR) HackRF device to sniff and decode GFSK-modulated (Gaussian Frequency Shift Keying) signals to interpret information being sent between a Z-wave Hub and a Z-Wave enabled device. The attack is used by sniffing the Traffic Analysis information, much like a Traffic Analysis Side-channel attack, however, this Side-channel attack is applied to the Wireless Sensor Network (WSN) communication sensor Z-wave protocol. |
|---|---|
| Bluetooth attacks | Common attacks to affect Bluetooth include *Bluejacking attacks*, *Bluebugging attacks*, *Car Whispering attacks, Bluesnarfing attacks* and *Denial-of-Service (DoS) attacks*. Other attacks include *Bluetooth Low Energy (BLE) attacks* such as *Passive Eavesdropping Man-in-the-middle (MiTM) attacks* and *Identity Tracking attacks*. These attacks could all facilitate potential Bluetooth equipped IoT device functionality manipulation, data theft, identity theft and the loss of IoT device functionality. Another significant IoT device Bluetooth vulnerability discovered was the *Bluetooth-based Timing attack*. |
| LoWPAN attacks | Low-Power Wireless Personal Area Networks (LoWPAN) are subject to potential IPv4 attacks such as *Denial-of-Service (DoS)* and *Man-in-the-middle (MiTM)* attacks. 6LoWPAN would potentially be subject to IPv6-based attacks such as *malicious Neighbour Discovery (ND) attacks*. Other LoWPAN attacks of significance are *RPL-Based 6LoWPAN Node Cloning attacks*, *RPL-Based LoWPAN Local Repair attacks, RPL-Based LoWPAN Increased & Decreased Rank attacks, RPL-Based Dropping Destination Advertisement Object (DDAO) attacks* and *RPL-Based 6LowPAN Routing attacks*. |
| GPS/GNSS Spoofing attacks | The process of a threat actor targeting the Global Positioning System (GPS) / Global Navigation Satellite Systems (GNSS) sensors of an IoT device such as Smartphone or Satellite Navigation System to direct vehicles or individuals to unsafe areas. it is a very simple attack to execute as GPS signals are often sent unencrypted and with the emergences of programmable radio platforms such as Universal Software Radio Peripheral (USRP), HackRF and bladeRF it has become much easier to build low-cost GPS/GNSS Spoofers. |
| GPS/GNSS Jamming attacks | An attacker, for as little as $15 can purchase a GPS/GNSS Jammer that will Jam a vehicles location-tracking services and affect the user's ability to track their vehicle if its stolen or use the road navigation systems built into the vehicle infotainment system, to get directions to different locations. This attack isn't mutually exclusive to Connected Autonomous Vehicles (CAVs). Any device that possesses a Global Positioning System (GPS)/Global Navigation Satellite System (GNSS) Sensor such as a Mobile Phone, Tablet, etc. will also be vulnerable to GPS Spoofing and Jamming attacks. |
| Sleep deprivation attacks | A dangerous IoT attack where the target maximizes device and/or sensor power consumption so that the lifetime of that device or sensor is minimized. This attack effectively causes wear and tear over the device's lifetime, and even though the attack is executed at the software level, it can still be considered a physical attack, as over time it causes physical damage to the device. |
| Energy depletion attacks | The main method of executing this attack is using a Spam DIS attack. This is where a malicious node generates multiple fictitious identities and sends a DIS request to increase the transmission process in the network and thus depletes the battery of the nodes. These attacks are regularly executed against the Routing Protocol for Low-Power and Lossy Networks (RPL) Routing system used in nodes operating within Wireless Sensor Networks (WSNs). |
| Sensor Identity attacks | The process of an attacker enumerating an IoT devices sensor to identify what the sensor does (disclosure of functionality) and using Machine Learning and Data Analysis techniques to probe the sensor for any vulnerabilities that an attacker can exploit through that specific sensor's vulnerabilities. |
| Replay attacks | A replay attack is a process whereby a threat actor retransmits authentication information in Cryptographically secure algorithms to deceive an interaction partner, thus allowing the attacker to gain control of the IoT Device. Replay attacks are passive in nature and often the frames of the packets have malicious or unusual timestamps to facilitate the likelihood of a Replay attack occurring. Replay attacks are most often used to steal electric vehicles, due to these vehicles using advanced signals to facilitate Remote Key Entry (RKE) into the vehicle and potentially facilitating vehicle theft. |
| Cellular Connection attacks | Attacks identified in the results consist of Denial-of-Service (DoS)-based attacks to cellular data network connections that IoT devices use. The attacks identified consist of *Third Generation (3G) Dedicated Channel (DCH) Starvation attacks*, *Fourth Generation/Long Term Evolution (4G/LTE) Roaming-based DoS attacks, 3G and 4G/LTE Overshadowing & Jamming attacks,* and *Fifth Generation Radio Resource Control (RRC) Replay attacks*. |
| Sensor Deception attacks | An attacker injects false malicious information directly into a devices sensor to deceive the sensors in order to falsify the sensors perceived environment to facilitate various malicious consequences. |
| Ultrasonic Side-channel attacks | An attacker sends malicious commands through a compromised browser on an IoT device to compromise the IoT devices ultrasonic sensors. |
| Acoustic Transduction attacks | An attacker targets the Inertial Sensors of consumer IoT devices by injecting malicious acoustics to trigger sensor measurement errors. |
| Node Capture attacks | An attacker captures a node present in a Wireless Sensor Network (WSN) and leaks secret information regarding authentication schemes and/or encryption designed to protect a device from future attacks and its data. |
| Fingerprint Recognition System (FRS) Spoofing attacks | An attacker deceives a biometric fingerprint reader to gain illicit access to a device or physical location guarded by this technology. |
| Acoustic Eavesdropping attacks | An attacker eavesdrops acoustics (audio/voice) information from any mobile devices speaker and/or microphones. |

*Table 5 Layer 1 - Physical Layer attacks*

**Layer 3** was the third most studied layer (23 studies), perhaps because it is used to manage a crucial aspect of Internet communications – routing. In fact, the selected studies included attacks such as node impersonation [42] [43] or sinkhole [44]. There are instances of DDoS attacks happening at layer 3 exploiting, for instance, the Internet Control Message Protocol (ICMP) [34], while some DDoS attacks occur at layers 4 and 5 as indicated previously. Some of the attacks presented at this level are also Side Channel Attacks (attacks that use alternative means to gain information rather than exploiting the protocol itself). Examples of these attacks involve electromagnetic [45], acoustic [46], timing information [47].

| Attack | Description |
|---|---|
| ICMP Flooding attacks | The process of flooding a single or all possible devices on a network with Internet Message Control Protocol (ICMP) packets to overwhelm the device or devices resources such as CPU, Memory, or Bandwidth to affect the network or devices availability subsequently resulting in a Denial-of-Service (DoS) attack. |
| Wormhole MiTM attacks | Wormhole attacks are facilitated by having two or more malicious nodes situated on two separate networks that connects two separate networks together. Wormhole attacks exploit networks by enabling (for example) the modification of data packets and the interception of sensitive and private information in an eavesdropping attack. |
| Packet Altering MiTM attacks | Packet Modification Attacks exist at both the control plane and the data plane. The control plane is considered where the routing of packets occurs, and the data plane is where the packets are redirected to the physical devices (MAC addresses). The control plane is considered Layer 3 of the TCP/IP stack, and the data plane is considered Layer 2 of the TCP/IP stack. This allows the attacker to launch Packet Drop, Extraneous Packet Generation Attacks, Packet Reordering, Packet Modification and Packet Delay Attacks. |
| Denial-of-Service (DoS) attacks | An attack that affects the service availability of network devices or services of legitimate users by flooding the network and/or specific device with useless traffic. Resources that are usually consumed on a network server or device are typically, CPU, Memory and Network Bandwidth. |
| Distributed-Denial-of-Service (DDoS) attacks | An attacker uses additional nodes to attack an IoT device to increase resource consumption to a greater level than that of a DoS attack, as this causes even greater latency. Latency is described as the amount of information the victim device receives and the greater this is, the higher the level of disruption. |
| Flooding attacks | An attack whereby an adversary uses various protocols such as TCP, UDP, HTTP or ICMP to overwhelm the device and/or network with packets of these protocols to the point that it affects network traffic and/or exhausts device and/or network resources such as CPU usage, Memory, or Network Bandwidth to the point that normal network operations are adversely affected or cannot be used at all. This effectively facilitates a DoS situation as it affects device and/or network availability. |
| HELLO Flooding attacks | HELLO packets are used for IoT devices and Wireless Sensor Networks (WSNs) to discover neighbour nodes near them. In this attack an attacker will use this technique but profusely use this to saturate the network, potentially causing sensor nodes to drain their energy reserves and facilitating a Denial-of-Service (DoS) attack. |
| Rushing attacks | A Wireless Mesh Network (WMN) attack. The attack works to disrupt the routing process between wireless nodes in a mesh network by exploiting the route discovery phase. A hostile node launching this attack, broadcasts the rushed Route Request (RREQ) message before any other intermediate node by ignoring the delay. This then increases the likelihood of the malicious node being included in the active routing path causing a flood of the data plane resulting in a Denial-of-Service (DoS) attack. |
| Routing attacks | The process of manipulating the routing process between the endpoint and the network gateway (Router). This can be done on the device directly or can be done on the router to affect the routing of packets off and on a network, to compromise the routing of the entire network and not one single device. Common Routing attacks include *Node Impersonation attacks, Sinkhole attacks*, *Gray Hole attacks, Black Hole attacks, Wormhole attacks*, *Routing Table Poisoning attacks*, *Byzantine attacks*, *Sybil attacks*, *Identity Forging attacks*, *Masquerading attacks* and *De-synchronization attacks*. |
| Gateway Bypass attacks | In this attack an adversary may bypass the gateway node and send commands directly to the function node. For safe data communications on and off a network (i.e. from Layer 2 to Layer 3 and from Layer 3 to Layers 7) it is vital to ensure that data goes through a trusted gateway over a malicious gateway if user privacy is to be maintained. The most common Layer 3 Gateway device is the Consumer Router. |
| Bit Flipping attacks | The network server receives a message from gateways, checks its integrity, and then sends the encrypted message to the application server that accepts it without any integrity verification. Thus, an intruder that sets up a man-in-the-middle (MiTM) attack between the network and the application servers, with a knowledge of message physical payload format, may sniff the message, modify some bits and send it to the application server. For example, in a smart home application, an intruder may set up a Bit Flipping attack so that there is a rise in the power consumption measured by an end device. The application server will get the consumed energy data without being capable to pick up the applied change. |
| Packet Replay attacks | An adversary intercepts and records data packets as they travel through the network to be transmitted later and subsequently, attempting to deceive the system into thinking the information is legitimate. |

*Table 6 Layer 3 - Network Layer attacks*

**Layer 6** was the fourth most studied layer with 16 studies identified. Attacks targeting Layer 6 mostly target data encryption methods through SSL/TLS Flooding attacks [48] a type of DoS attack that prevents Secure Socket Layer (SSL) or Transport Layer Security Encryption, Side-channel attacks [49] by leaking cryptographic keys by analysing electromagnetic [45], traffic [50] or timing behaviour [47] or potential future attacks through Quantum Computing attacks [51] to break secure encryption methods to expose user's private and/or sensitive information or data.

| Attack | Description |
|---|---|
| Side-channel attacks | These attacks analyse the information of an electronic system available through side-channels, such as the Power Consumption, the electromagnetic (EM) emanation, or the timing behaviour of the system. Types of Side-Channel attacks include, *Traffic Analysis attacks*, *Cache Timing attacks*, *Electromagnetic Analysis attacks*, *Simple Power Analysis (SPA) attacks*, *Differential Power Analysis (DPA) attacks*, Differential *Fault Analysis attacks* and *Acoustic Analysis attacks*. |
| SSL/TLS Flooding attacks | This attack uses the need to expend computing power of the server when building a secure Transport Layer Security (TLS) or Secure Socket Layer (SSL) connection needed to provide end-to-end encryption for confidential, sensitive information sent from the IoT Device to the Cloud. In this attack the adversary loads the server's resources beyond its limits and shutting it down during TLS negotiation by sending many packets to the server or constantly asking to renegotiate the connection. An SSL/TLS Flooding attack mainly consumes a web server's CPU resources to affect sensitive information being sent to the server by individuals and thus affecting data security and confidentiality. |
| Forward Secrecy attacks | The process of an adversary attempting to steal the secret keys and session keys, without the use of a Side-Channel Analysis (SCA)-based attack method, during the communication of data between an end-device and its destination for the purposes of decrypting the packet and gaining illegal access to the information. |
| Quantum Computing attacks | While current cryptographic systems are capable of encrypting data transmitted between devices and the cloud using very large keys and advanced encryption algorithms such as elliptic curve cryptography (ECC) there is new computing power on the horizon. This is known as quantum computing which makes use of quantum entangled states measured in qubits. Quantum computing has already been shown to have enormous impact on currently infeasible |

|  | calculation problems. The major risk is that once quantum computing is more advanced, currently popular public-key cryptography algorithms such as RSA and elliptic curve cryptography will be easily defeated. |
| --- | --- |
| Message Tampering attacks | An attacker modifies the content of a message that is being transmitted to make an unauthorized impact by re-ordering or delaying it. |

*Table 7 Layer 6 - Presentation Layer attacks*

**Layer 2** was the fifth most studied layer, with 12 studies identified. Attacks of this nature usually include those that target a user's data, such as Man-in-the-middle (MiTM) attacks [52], ARP Poisoning [53], Spoofing [54] or Eavesdropping [55].

| Attack | Description |
| --- | --- |
| Man-in-the-middle (MiTM) attacks | An attack that allows an offender to sniff or capture data using attacks such as but not limited to *ARP Poisoning* and *Spoofing attacks*, in conjunction with, packet capturing tools such as, Wireshark or TCPdump to capture and then receive private sensitive information. |
| ARP Poisoning attacks | ARP Poisoning or Spoofing is the process of tricking the Layer 2 Switches ARP table to add another ARP entry into the table that maps the logical IP (Internet Protocol) address (Layer 3) to the Media Access Control (MAC) address at Layer 2 to a malicious access point controlled by the attacker, alongside a IP-to-MAC ARP address mapping to the intended destination. This effectively puts the attacker in the middle of the conversation between the destination and the sender of the packets and is considered the most common MiTM attack. |
| Spoofing attacks | The manipulation of fake identities to compromise the effectiveness of the IoT device by forging many identities to act as legal nodes. This then allows an attacker to intercept sensitive information from an IoT device present on a Local Area Network (LAN). |
| Eavesdropping attacks | Eavesdropping is a type of MiTM attack that consists of capturing and deciphering sensitive information that is stolen from the victim while the MiTM attack is taking place from *ARP Poisoning* or *Spoofing attacks*. |
| Packet Altering MiTM attacks | Packet Modification Attacks exist at both the control plane and the data plane. The control plane is considered where the routing of packets occurs, and the data plane is where the packets are redirected to the physical devices (MAC addresses). The control plane is considered Layer 3 of the TCP/IP stack, and the data plane is considered Layer 2 of the TCP/IP stack. This allows the attacker to launch Packet Drop, Extraneous Packet Generation Attacks, Packet Reordering, Packet Modification and Packet Delay Attacks. |

*Table 8 Layer 2 - Data Link Layer attacks*

**Layer 4** was the sixth most studied layer with 7 studies identified. This layer largely focussed on Profiling a victim's device through Reconnaissance attacks utilizing tools such as NMAP [56] and Shodan.io [57]. Additionally, some DoS-based Flooding attacks target this Layer through attacks such as TCP-SYN Flooding [32] and UDP Flooding [33] attacks.

| Attack | Description |
| --- | --- |
| TCP-SYN Flooding attacks | This attack usually targets and exploits the Transmission Control Protocol (TCP) Three-Way Handshake with the objective of making any server unreachable for legitimate access. The attack initiates repeated attempts at step two of the handshake known as SYN (Synchronization) messages to the victim without completing step three in the Three-Way Handshake process to facilitate a Denial-of-Service (DoS) attack. |
| UDP Flooding attacks | Like TCP-SYN Flooding attacks, however, instead of the attacker abusing the TCP protocol they use the User Datagram Protocol (UDP) instead. In this attack the victim is flooded by a burst of UDP packets to specific UDP services/ports that a device may be broadcasting to facilitate a Denial-of-Service (DoS) attack. |
| Flooding attacks | An attack whereby an adversary uses various protocols such as TCP, UDP, HTTP or ICMP to overwhelm the device and/or network with packets of these protocols to the point that it affects network traffic and/or exhausts device and/or network resources such as CPU usage, Memory, or Network Bandwidth to the point that normal network operations are adversely affected or cannot be used at all. This effectively facilitates a DoS situation as it affects device and/or network availability. |
| Network Mapper (NMAP) | NMAP is primarily used to fingerprint local and remote networks for endpoints listening on active TCP (Transmission Control Protocol) or UDP (User Datagram Protocol) ports to then allow an attacker to progress on to the exploitation of these services. It is also used to fingerprint applications and Operating Systems of devices, check devices for Critical Security Vulnerabilities (CVE's), Automate the pinging of hosts to see if they are alive and ready for |
| | Exploitation and even comes with a Scripting Engine, known as NMAP Scripting Engine (NSE), that allows hackers and penetration testers to write scripts or use the many built in Scripts to test systems for Vulnerabilities and Network Weaknesses. |
| Shodan.io | Shodan.io is a global website search engine for all publicly exposed consumer IoT devices. The Shodan API uses three trap categories to scan the internet for vulnerable exposed IoT devices. Trap categories are also known as device Honeypots. They allow devices that exist on the internet to be obtained and analysed within the Honeypot and are used to provide detailed information about devices and associated vulnerabilities. These traps are Dark Trap, White Trap and Red Trap. |

*Table 9 Layer 4 - Transport Layer attacks*

**Layer 5** was the least studied layer with 3 studies identified. These attacks consist of attacks that target the underlying technologies such as SOCKS4 and SOCKS5 socket protocols responsible for establishing a session or proxy. As proxies allow the attacker the ability to obfuscate their actions attackers often use proxies as a method of repudiation [58]. Whereas Device Hijacking attacks [59] allow an attacker to hijack the session entirely.

| Attack | Description |
|---|---|
| Repudiation attacks | The process of an attacker carrying out a specific attack and then using various obfuscation techniques such as Proxying by establishing a malicious protocol socket connection or the removal of a Connection/Malicious Node ID, so that the victim cannot trace the attacker's identity at the source and subsequently cannot alert the relevant authorities with this vital information. Non-repudiation means the opposite, where the victim takes actions to disallow the perpetrator from repudiating themselves. |
| Device Hijacking attacks | In this attack the adversary hijacks and gains control of the device through its established session. These types of attacks are difficult to detect because the attacker does not change the basic functionality of the device. Moreover, the adversary only manipulates one device to re-infect all smart devices in the home to paralyze the network. |

*Table 10 Layer 5 - Session Layer attacks*

### 4.2.3 Hierarchy of Evidence categorization

In this section we examine the types of methodologies researchers employed in the identified studies. Laboratory based studies were perhaps not surprisingly the most common amongst the literature (44.27%), while studies that involved the simulation of devices in a computational environment accounted for 20.16% of those reviewed. As such, around 65% of the papers employed some form of experimental design. Just over one-quarter of the papers (28.06%) fall under the expert speculative category, while the remaining 7.52% was split among author speculative (5.14%), where authors gave their own point of view), user speculative (0.40%, asking non-expert users opinions and information). Only 1.98% of the studies reviewed were conducted in real-world settings. It is not surprising that real-world research was so under-represented as there are ethical implications to exploiting devices in the wild.

### 4.3 Recommended mitigation strategies

296 papers were identified that discussed mitigations for cyber-attacks against consumer IoT devices. These are summarised in Figure 4. The countermeasures discussed were mainly designed to operate at the network level (162 out of 296), or focused on incident management software (SIEM, 91 out of 296). Fewer papers focused on device (25), cloud (11), or application specific mitigations (7). As before, we evaluate how the mitigations fit into the OSI stack (162 papers).

### 4.3.1 Mitigations categories

The remaining works were divided among Device attack mitigations which include Connected Autonomous Vehicles (CAVs) [60], Drones [61] and Routers [62].

Application attack mitigations consist of device attack mitigations that target their specific application scenarios for Multimedia Systems [63], smart TVs [64] and smartphones [65].

Cloud attack mitigations consist of mitigation strategies to devices that require exclusive use of the cloud to function, with this device being Smart Gas and Electric Meters [66] only.

Another category of mitigation strategy discussed were Security Information and Event Monitoring (SIEM)-based countermeasures which look at mitigations that are not specifically designed for IoT devices but could be applied to them. Studies that discussed such approaches (91 studies) considered mitigation strategies such as Intrusion Detection Systems (IDS) [67], Authentication mechanisms [68], Blockchain-based countermeasures [69], Access Control measures [70] and Best Practice solutions [69].

### 4.3.2 Mitigations categorised through the OSI stack

As was the case for attacks, the Network category mitigations most discussed, focused on Layer 7 (N=56). Layer 7 attack mitigations consist of largely malware attack mitigations [71]. However, Hyper-text Transfer Protocol (HTTP) exists at Layer 7 of the OSI stack and as such mitigations to DoS attacks through HTTP Flooding [72] have also been identified by researchers in the literature. DNS attack mitigations [73] were also present in the literature and are categorized under Layer 7 of the OSI stack. Approaches include a **network monitor** [74] [75], an **edge assisted anomaly detection framework** [71], a **network firewall** [76], an **AI-powered honeypot for enhanced detection of IoT Botnets** [77], a **lightweight approach using Machine Learning (ML) algorithms to detect DDoS attacks caused by Botnets** [78], **Anomaly-based, Signature-based, Network-based, DNS-based, Host-based and Blockchain-based detection methods** [79], a **Network Intrusion Detection System (NIDS) using Gradient Boost Decision Trees (GBDT)-based XGBoost & Rough Set Theory** [80] and the use of **Stacked Recurrent Neural Networks (SRNNs) for the detection of Botnets** [81] to mitigate Botnet malware. Typically, the most common ways to prevent a Ransomware attack is to **update consumer IoT device firmware**, **keep track of all IoT devices on the network**, **install a network monitoring system, back up data, update anti-malware programs, refuse to pay the ransom, education, disallow admin privileges** and **flag any suspicious** traffic [82] [36]. Additionally, an **early detection method using Hardware Performance Counters (HPC)** [83] is proposed to detect Ransomware during its early stages of execution to prevent the attack from executing. Finally, researchers used a **deep learning system for detecting ransomware** [84] in edge computing devices. DNS attack mitigations consist of, **DNSSEC** [73] to mitigate DNS poisoning attacks, an **Auto Configuration-Based Enhanced and Secure Domain Naming Service for Internet Protocol (IP) Version 6 (IPv6)-Based Domain Name System (DNS)** [85] to prevent DNS Water Torture attacks and **sFlow** [86].

The second most studied layer for attack mitigations was Layer 3 (N=53) of the OSI Model. This contrasts with Layer 1 being the most studied Layer of the OSI Model for the attack results. These consist of network communication attack mitigations such as those that address masquerading attacks [87], Black hole attacks [88], Ad-hoc On-demand Distance Vector (AODV) [89] Routing protocols for Mobile Ad-hoc Networks (MANETs) and Vehicular Ad-hoc Networks (VANETs) and Optimal Link State Routing (OLSR) [90]. Other mitigation strategies discussed included those that address DoS attacks regarding Internet Control Message Protocol (ICMP) flooding attacks [91], Cryptographic attack mitigations for Side-channel Analysis (SCA) attacks [92] and Data attack mitigations for eavesdropping through Internet Protocol Version 6 (IPv6)-based routing [93]. Approaches included a **DoS-free Optimal Link State Routing (DFOLSR) protocol** [90] to protect VANET and MANET networks from DoS-based routing attacks, **a secure and lightweight authentication with key agreement scheme for smart wearables systems** [94], a **novel smart home authentication protocol known as LRP-SHAP** [95], **a certificate-based authentication scheme for smart homes** [96] to detect impersonation attacks, **a lightweight and robust security protocol for smart home environments known as LR-AKAP** [97], a security scheme known as **Enhanced Secure Device Authentication (ESDA) scheme** [98] to protect against node impersonation attacks, **a novel decentralized security architecture that monitors the routing of packets between the edge network and the IoT devices to mitigate sybil attacks in RPL-based (routing protocol for low-power and lossy Networks)** [99], a **large-scale smart home identity System using hierarchical identity-based encryption for the mitigation of masquerading attacks** [87], black hole attack mitigation strategies include a **prevention mechanism using the roulette wheel selection technique** [88], **an anomaly-based Support Vector Machine (SVM) Intrusion Detection System (IDS)** [100], an **enhancement of the AODV routing protocol in use in a MANET network** [89], a **mitigation strategy for Network-on-Chip (NoC)-based routers** [101], an **enhanced performance AODV route establishment mechanism** [102] and **sequence number-based detection, protection using Cryptography and opinion from other nodes in the MANET or VANET network-based mitigation strategies** [103].

Layer 1 attack mitigations were the third most frequently discussed (N=27 studies). These included mitigations for Wi-Fi jamming attacks [104], Wi-Fi evil twin attacks [105], Radio Frequency Identifier attacks [106] and Wireless Sensor Network (WSN) attacks [107]. Approaches included monitoring the **Received Signal Strength (RSS)** [108] to mitigate Wireless Eavesdropping attacks. While Wi-Fi Evil Twin attack mitigations include **TrustedAP** [105], **wireless channel RSSI** [109], monitoring of **Packet Delivery Ratio (PDR) and to mitigate Wi-Fi Jamming/De-authentication attacks; Bad Packet Ratio (BPR), Energy Consumption Amount (ECA) and Signal-to-Noise Ratio (SNR)** [104] were proposed. Radio Frequency Identifier (RFID)-based countermeasures included a **lightweight key authentication mechanism known as LW-AKA** [110], a **trust-based RFID Authentication System (TRAS)** [106] and a protocol known as **RUND** [111] to defend against DoS, replay and timing attacks. A new security protocol known as **WZ-lcp (W2-ZigBee Low-Cost Protocol)** [112] was developed to protect ZigBee devices from common attacks. A **mutual authentication scheme with user anonymity for cyber-physical and IoT devices using Burrows-Abadi-Needham (BAN) logic** [113] was developed to mitigate NFC replay attacks. A **secure IoT system that uses Bluetooth** [114] was developed to ensure all packets are sent over Bluetooth securely. An **Intrusion Detection Framework for the IoT empowered 6LoWPAN** **protocol** [107], a **secure and enhanced authentication and key establishment scheme for 6LoWPAN known as EAKES6Lo** [115] and an **Information Centric Networking (ICN) system** [116] to mitigate 6LoWPAN attacks. A low-cost GPS spoofing detector design was recommended using a **hardware oscillator, anomaly detection techniques, hardware-oriented security and trust-based** [117] to mitigate GPS spoofing attacks. To mitigate against the Energy Depletion attack-based SPAM DODAG Information Solicitation (DIS) attack, a system known as **DIS Spam Attack Mitigation (DISAM)** [118] was proposed.

Layer 6 attack mitigations were the fourth most frequently discussed (N=13). These included encryption mitigation strategies such as Secure Socket Layer/Transport Layer Security (SSL/TLS) flooding [48] and Side-Channel attack [49] [119] [120] solutions. Approaches include an **implementation of the Datagram Transport Layer Security (DTLS) security scheme** [121] and **Threat-TLS** [122] for SSL/TLS Flooding attacks and a **Ring Oscillation (RO) method** [92] and a **behaviour monitoring & similarity comparison technique** [123] to mitigate common Side-Channel attacks. A **traffic obfuscation for smart home Local Area Networks (LAN)** [124] for the prevention of traffic analysis attacks. **Edge-CaSCADe** [125] to mitigate Timing-based Side-Channel attacks. An **improvement of the Saber SCA attack mitigation technique using a High-performance Masking Design Approach utilizing hash functions with cantered binomial sampling & masking conversions** [126], a **novel dynamic instruction scheduler known as PARADISE to mitigate Simple Power Analysis (SPA) attacks** [127] and a system to mitigate **acoustic adversarial attacks on microphone-equipped smart home devices using Deep Neural Networks (DNNs) and Machine Learning (ML)** [128]**.**

Layer 2 attack mitigations were the fifth most frequently discussed (N=12). Layer 2 attacks include eavesdropping attack mitigations [129], Man-in-the-middle (MiTM) attack mitigations [130], IPv6-based eavesdropping attack mitigations [131], Address Resolution Protocol (ARP) Poisoning attack mitigations [132] and Spoofing attack mitigations [133]. Approaches include Elliptic curve cryptography (ECC) solutions [134]. Mitigations to IPv6 attacks on Gateway Routers would be to use **Authorised Delegation Discovery (ADD) mechanism of SEND**, the **Trusted Router Discovery Protocol (TRDP)** and **Router Advertisement (RA) Guard**. However, these mitigation strategies can be susceptible to collision attacks and bootstrapping issues [131]. Potential alternative solutions are the **SecMac-SRD mechanism** [131] and **fingerprint-based RA Guard (FibRA-Guard)** [93]. ARP Poisoning attack mitigation strategies consist of a **Software Defined Networking (SDN)-based framework for enforcing network static and dynamic access control** [132] and **port security** [135]. While Spoofing attack mitigations include **a cutting-edge Intrusion Detection System (IDS) / Intrusion Prevention System (IPS) using Artificial Intelligence (AI)** [133]. Mitigations for eavesdropping include a **lightweight encryption scheme for smart home environments** [136], a **lightweight enhanced collaborative key management system** [137], **lightweight key agreement and authentication protocol for smart homes using exclusive or (XOR) and hashing operations** [129]**,** a **man-in-the-middle (MiTM) attack resistant secret key generation scheme via channel randomization** [130]**.** Packet Altering MiTM attack approaches include an **agile approach for data protection known as Data Protection Fortification (DPF)** [138]. Data Exfiltration attack approaches include the use of an **automated data exfiltration detection system utilizing the MITRE ATT&CK Framework** [139].

Layer 4 mitigations were the sixth most frequently discussed (N=7). Layer 4 attacks include two Flooding attacks [140] [141]. These are TCP-SYN [32] and UDP Flooding attacks [33]. Additionally, Reconnaissance attacks occur at this layer, mainly using the Network Mapper (NMAP) [56] and Shodan.io [57] [142] tools. Approaches include the use of an **Intrusion Detection System (IDS) using the iptables firewall and access control system** [143] to filter traffic that arouses suspicion in the form of repetitive packets for Flooding attacks. TCP-SYN (Half-Open) flooding attack mitigation strategies include the use of a **real-time application DDoS detection system known as ForChaos** [144]. While UDP flooding mitigations include the use of a **UDP flooding-based DDoS attack detection method** [145] based on protocol specific traffic features.

Layer 5 mitigations were the least studied with no studies discussing mitigation strategies.

### 4.3.3 Hierarchy of Evidence categorization

In this section we summarise the methodologies researchers employed in the studies reviewed. Experimental studies that were either laboratory based (60.47%) or used simulations (21.96%) were the most common. Several papers (12.84%) fall under the expert speculative category while the remaining 4.73% was split among author speculative (3.38%), where authors gave their own point of view, and real-world settings (1.35%). The user speculative category had no published works. The relative absence of real-world studies highlights the need for lab-based studies to emulate real-world conditions as closely as possible.

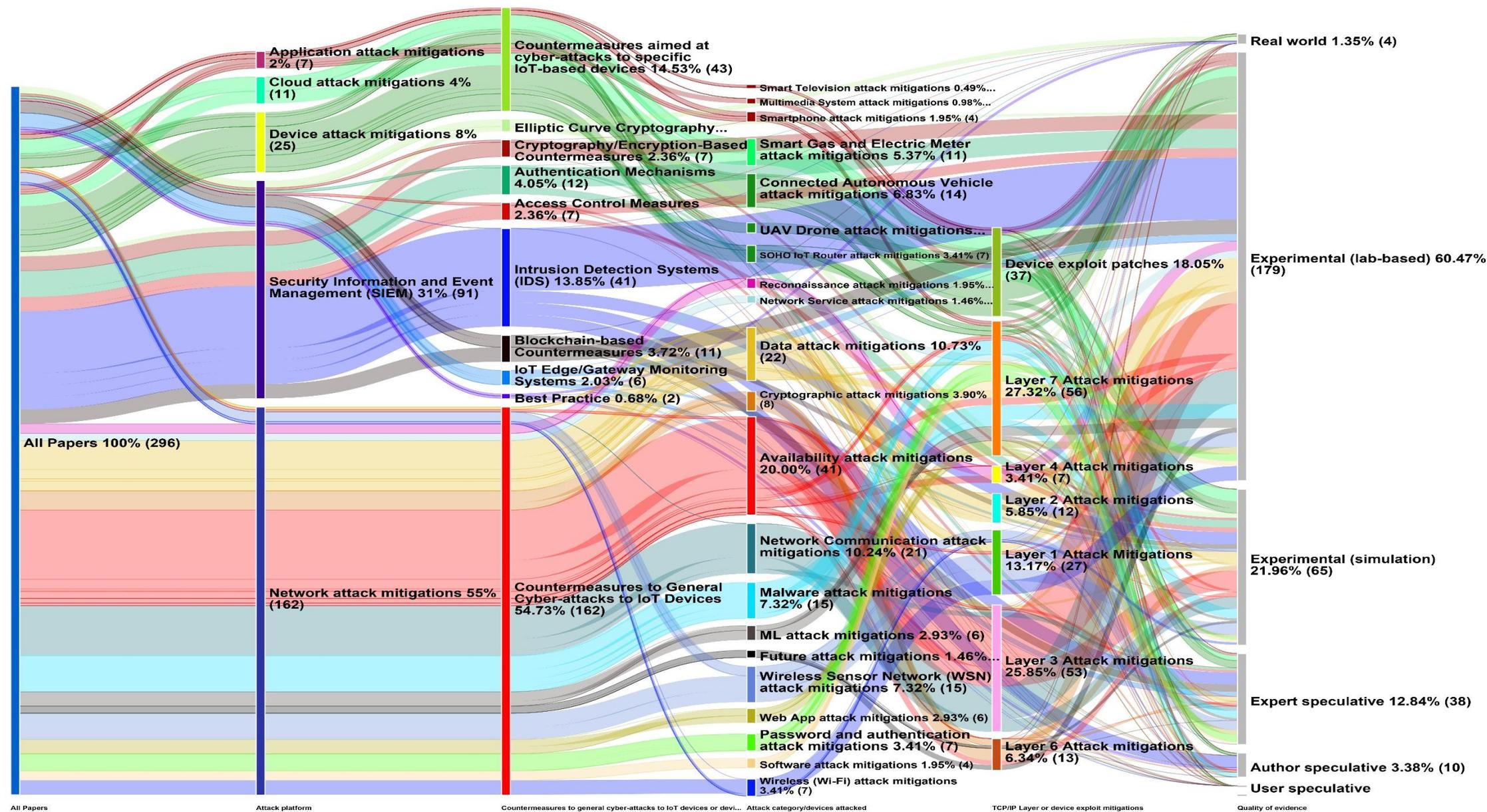

*Figure 4 Sankey Diagram to illustrate security countermeasures.*

## 4.4 Crimes enabled by attacks against consumer IoT devices

Many security attacks against IoT have the potential to facilitate crime. Many crimes were identified in this review including residential burglary, money laundering, identity theft, fraud, spying and crypto jacking. Below, we discuss

A similar systematic review of crimes facilitated by consumer IoT devices was conducted by Blythe and Johnson (2021) [4]. In that review the authors identified mechanisms (i.e. things that can be done to facilitate crimes) and the actual harms/cybercrime that they facilitate. However, the searches conducted for that review were completed in 2017 and the security landscape has moved on significantly since then. As such, this review presents additional mechanisms and crimes (See Table 11 and Table 12, respectively) that are now facilitated by the recent evolving security landscape for consumer IoT devices. These additional mechanisms consist of hacker training [146] where the hackers train additional recruits to carry out attacks to commit cybercrime, hacker recruiting [146] where a hacker recruits either already skilled hackers or non-skilled hackers to then train as part of the hacker recruiting mechanism to commit cybercrimes, criminal marketplace/criminal reputation/value of criminal activity [146] where attackers increase their reputation as part of a potential CaaS model (See Below) to improve their reputation to non-skilled criminals. Additional crimes consist of fraud [147], hardware damage [148] by using software to cause damage to the device so that it no longer boots bricking the device, annoyance [149], money laundering [146] where an attacker can utilize attacks such as Botnets to mine cryptocurrencies using consumer IoT devices facilitating a Cryptojacking crime [150] and Intimate Partner Abuse (IPA)/Domestic Abuse (DA) [151] where individuals participated in a profile survey and asked to indicate whether they believe that their smart devices could be used for the purpose of allowing a coercive controller to take control of their equipment to cause malicious consequences to the victim.

Over the last few years, the cybercrime ecosystem has evolved with many criminals offering Cybercrime-as-a-Service. With the cybercrime ecosystem evolving to include attackers providing their technological expertise to criminals lacking a technical background (a process known as Proxying).These crime type services consist of many different variations of Cybercrime-as-a-service (CaaS) [146] models. CaaS models allow technically illiterate criminals to utilise attacks and exploits from skilled hackers as a service. Some of the identified CaaS models include Exploit-as-a-Service (EaaS) or Exploit-Package-as-a-Service (EPaaS), Deception-as-a-Service (DaaS), Payload-as-a-Service (PaaS), Obfuscate-as-a-Service (OBaaS), Money-Laundering-as-a-Service (MLaaS and Hacker-Recruiting-as-a-Service (HRaaS). The full list of CaaS models identified by Huang (2018) [146] will be provided in Table H3b – Cybercrime Mechanisms.

| Mechanism | Definition | Real World | Experimental (Lab-based) | Experimental (Simulation) | Expert Speculative | Author Speculative | User Speculative |
|---|---|---|---|---|---|---|---|
| Digital Gains Data theft Exposing personal user data | An attacker can obtain information contained in a victim's system, including sensitive information such as personal profiles, accounts, and intellectual property. | | [152] [153] [154] [109] [155] [156] [157] | | [158] [146] [159] [160] | [161] [162] [163] [164] [165] [166] [167] [4] | |
| Profiling | Attackers maliciously monitor a user's activity to gauge level of activities being conducted (e.g. walking, running, cycling, etc.) to profile their behaviour | | [168] [169] [170] [171] [172] [173] [174] [175] [176] [177] [178] [179] [180] [181] | [182] | | [183] [184] [185] [186] [187] [188] [189] [4] | |
| Physical Access Control | Attacker misuses devices linked to physical access in the home. | | [154] [190] [191] [149] [192] | | | [4] | |
| Manipulation of device functionality | Attacker remotely controls and manipulates the device. For example, using actuators on household robots to cause damage to household property. | | [154] [191] [192] [193] [194] [195] | | [5] | [4] | |
| Control Audio/Visual Outputs | Use of audio/visual outputs of IoT devices to control what the user hears/sees. | | [156] [195] [196] [197] [198] [199] | | | [4] | |
| Suppress safety-critical monitoring capabilities | Malicious control or suppression of safety-critical monitoring devices (e.g. fire alarms). | | [154] | | | [187] [188] [189] [190] [191] [149] [192] [193] [194] [195] [196] [197] [197] [198] [200] [201] [4] | |
| Service Availability and/or Restriction | Connected devices are linked to services in the home including critical (e.g. physical access, heating) and less critical (e.g. internet access) ones. Exploitation can lead to denial-of-service for consumers or censorship of certain product functions. | | | | | [161] [184] [186] [194] [202] [203] [204] [205] [206] [4] | |
| Monitoring/Surveillance | Exploitation of consumer IoT devices may allow attackers to listen and monitor user activities. | | [157] [178] [195] [207] | | [158] | [161] [193] [200] [208] [209] [4] | [167] |
| Gateway to further attacks | Once devices are exploited, attackers may use the device or information gained from it to launch additional attacks. For example, using a device as part of a Botnet to launch | | [210] | | [158] | [162] [183] [184] [196] [208] [202] [4] | |
| | DDoS attacks, or using personal information for targeted password guessing. | | | | | | |
| Hacker Training | Since most hackers are novices, part of the value-added activity for the hacker community is training the novices. | | | | [146] | | |
| Hacker Recruiting | To grow the hacker community, recruiting is an important activity for the cybercrime ecosystem. To achieve this goal, many tutorials are available to reduce the barriers for novices to join the hacker community and benefit from the cyber-attack. | [146] | | | | | |
| Criminal Marketplace Criminal Reputation Value of Criminal Activity | A marketplace for attackers to trade the digital gains is the principal way for attackers to realize the benefit from successful cyber-attacks. Criminals rely on a hacker's reputation and potential value when exploring the marketplace. | [146] | | | | | |

*Table 11 Cybercrime Mechanisms*

| Cybercrime | Definition | Real World | Experimental (Lab-based) | Experimental (Simulation) | Expert Speculative | Author Speculative | User Speculative |
|---|---|---|---|---|---|---|---|
| Energy Theft | The attacker uses IoT-enabled smart meters to manipulate the electricity consumption measured by the smart meter. The attacker could reduce the energy usage reported by other smart meters in the community. As a result, victim customers receive elevated electricity bills while the aggregate bill for all customers in the community remains the same and the price for the attacker is reduced. | | [194] [211]<br><br>[212] [213] | [214] [31] [215] [216] [217] [218] [219] [220] [221] | | [184] [185] [222] [161] [223] [187] [224] [167] [225] [4] | |
| Residential Burglary/Burglary/Physical Theft | Due to the increase of cyber-enabled home environments increasing, this increases the possibility of an attacker using computers to increase the scale or reach using computers. Examples could include the tampering of surveillance recordings of residential properties or stopping surveillance from performing its intended function allowing a perpetrator to enter the property unannounced, with no evidence linking the perpetrator to the crime. | | [190] [191] [154] [149] [226] | | [158] [227] | [222] [225] [177] [190] [149] [226] [163] [183] [178] [188] [209] [4] | |
| Sex-crimes | Use of consumer IoT devices to facilitate sex-related crimes such as stealing sex-related videos, sexual assault, obscenity, exhibitionism, and voyeurism. | | | | [158] [228] | [4] | |
| Political | Exploiting consumer IoT devices for political gains (e.g. political subjugation and control, and propaganda). | | | | [158] [228] | [4] | |
| Identity Theft | An attacker steals sensitive information stored on an IoT device or associated Cloud Server, pertaining to the victim, which then an attacker can use for malicious illegal purposes. | | [147] [165]. | | [228] | [184] [183] [209] [4] | |
| Harm to individuals/Homicide | Causing physical or mental harm to individuals including vulnerable groups (e.g. children and older adults) that may be susceptible to nefarious influence. For example, targeting devices with heating capabilities to cause a fire in the home. | | [192] | | | [187] [224] [186] [193] [223] [188] [229] [4] | |
| Misinformation | Use of IoT devices to give false or inaccurate information (e.g. false fire alarms) or to manipulate pre-existing information present within IoT devices. | | | | | [149] [226] [165] [4] | |
| Direct Monetary Gains Financial Loss Crypto Jacking | The attacker can make a profit and benefit by monetizing the victim's loss for themselves. Typical scenarios involve the attacker drawing funds from a victims account from information obtained from sensitive information on the devices attacked. | [146] [230] [231]. | [150] | | [158] | [161] [224] [186] [154] [164] [203] [4] | [150] |

| | | | | | | |
|---|---|---|---|---|---|---|
| Unsolicited advertising | Use of information from IoT for targeted advertising and marketing. | | | | | [183] [4] [209] |
| Blackmail/Extortion | Use of information gained from IoT devices to blackmail individuals. | | | | [158] [228] | [161] [195] [4] |
| Vandalism/Criminal Damage | Damage to physical property or household objects arising from exploited devices with actuators. | | | | | [149] [226] [193] [4] |
| Illicit affective response | Use of information gained from IoT devices to cause embarrassment, annoyance, or damage reputations. | | | | | [149] [228] [193] [4] |
| Discrimination | Misuse of information from IoT devices (e.g. beliefs, health information) to discriminate against individuals. | | | | | [167] [183] [4] |
| Stalking Spying | The unauthorized monitoring of a victim through Smart Home devices (e.g. Smart Cameras, Smart Doorbells, etc.) to monitor a victim through surveillance or sensitive information data stored on IoT devices for malicious purposes | | | | | [183] [209] [147] [4] |
| Money Laundering | A traditional activity for underground crime, to make illegally gained proceeds appear legal. | | | | [146] | |
| Fraud | The process of using cyber-attacks on IoT smart home devices to facilitate the illicit appropriation of money, cryptocurrency or other benefits using malicious or deceptive means. | [232] | | | | [147] |
| Hardware Damage | Certain security attacks may cause physical hardware damage to a device and/or its components. An example of this was the Chernobyl malware that allowed the attacker to reflash the BIOS Chip, corrupting the bootstrapping program required to initialize the system. | | | | [148] | |
| Annoyance | An attacker could spoof physical events on a smart device to facilitate random or constant sounding of the alarm. This could cause annoyance and frustration to the consumer. | | | | | [149] |
| Psychological Gains | The attacker who carries out attacks seeking the inherent satisfaction of success for the fun or challenge of the process gains psychological benefits from an attack. In this case the attack is perceived as merely a test of hacking skills, | | | | [146] | |

| Intimate Partner Abuse (IPA)/Domestic Abuse (DA) | A type of crime that perpetrators inflict on the victim through various methods such as Coercive Control, Gaslighting, Sexual Violence and Continuous Physical Harm/assault to the victim. | [151]. | | | | |

*Table 12 Cybercrimes*

## 4. DISCUSSION

The results of the systematic review on crimes facilitated by security vulnerabilities to consumer IoT devices raised several topics of interest. These were the actual attacks discussed in the literature, the mitigation strategies of such attacks discussed in the literature, the crimes identified from the results of the systematic review and potential limitations of the methodology used in this study. These are all discussed in the following subsections.

**Attacks discussed in the literature**

Few studies were conducted in the real-world with most taking place in experimental laboratories or computer simulated environments. This is expected due to ethical reasons and the risks associated with conducting attacks in real-world scenarios. For example, a researcher might unintentionally damage a consumer's device and/or collect data about the owner or the participant. Moreover, attacks against (say) connected autonomous vehicles or drones could result in collisions or injury. Doing so might also violate various laws regarding data protection, criminal damage (or other offences) if the attack damages the device the attack was tested on. However, to properly gauge how attacks, mitigation strategies and resultant crimes might be impacted to real consumers, more studies need to be conducted in a real-world setting.

Surveys in the Expert (Speculative) category are not conducted as members of the public are likely unaware of the technicalities of cyber security and their attacks to consumer IoT devices. However, some literature reviews were identified, where the researcher reviews already published works on security attacks to consumer IoT devices and survey's them in their own study. Another methodology that was rather popular among the identified papers was surveying experts to effectively understand the threat scenarios. While this methodology may not necessarily present new attack vectors, it gives interesting insights without incurring in the ethical or legal hurdles mentioned. These hurdles affect real world experimentation that, in fact, was rarely used as methodology. This strategy of using experimental or simulated scenarios should be continued by researchers as conducting this in an isolated, experimental setting mitigates the ethical issues already mentioned.

Potential future attacks that could be facilitated such as wide-scale power blackout attacks to smart gas and electric systems were surprisingly not discussed. This could cause issues in power critical environments (i.e. hospitals, emergency services, etc.) and as such further research would be needed on these topics. Application attacks towards Smart TVs mention the possibility of attacks to cameras present on some newer TVs but do not mention the possibility that an attacker could use this to spy on victims. This same scenario occurred with Smart Multimedia systems where researchers discuss attacks that enable an attacker to hijack the voice command control matrix to pivot onto a consumer's network and take control of other devices but do not mention the possibility that consumers could be spied on using their Alexa, or Apple HomePod, for example. Additionally, in the case of application attacks, several popular consumer IoT devices were omitted from discussions. These included Smart Doorbells, Tablets, Smart Watches, Smart Fridges, Smart

Cookers, and Smart lightbulbs. Attacks aimed towards CAVs included sensor attacks to the LiDAR and Radar systems, however, again researchers do not explicitly discuss the causal link between specific attacks and crime. Attacks to CAVs through LiDAR or Radar sensors could be used by hostile nation states to commit terrorism remotely, mass murder, physical assault or criminal damage. However, again this implication is not discussed in the selected works identified in the Systematic Review by the researchers. An overall observation is that researchers mainly focus on conducting attacks against the components of devices without discussing the potential impact that these attacks could have on society. Drones could also be used by hostile nation states to commit acts of terrorism, but this type of offending was not mentioned in any of the papers. Instead, it is common for researchers to note that perpetrators could take remote control of devices (e.g. drones) but to not then say how they can be misused for criminal purposes. The importance of researchers outlining these particular impact scenarios and possible facilitating offences, would allow the researchers to give far greater weight to their claims when demonstrating these attacks and would allow them to more properly convey what the resultant outcomes to victims would be, if these attacks were attempted in a real-world setting.

In regard to legislative bodies around cybersecurity and cybercrime prevention to consumer IoT devices, there is no specific IoT device legislation that solely regulates consumer IoT devices only. This is with the exception of Internet-of-Vehicles (IoV) devices. With IoV devices they are governed by the Connected Autonomous Vehicles Act. However, consumer IoT devices as a whole fall under legislation by the British for all connected devices. For this reason, a specific consumer IoT device legislation that focusses solely on consumer IoT devices, with the rapid increase of potential attacks to these devices, should be developed.

**Mitigations discussed in the literature**

Overall security attack mitigation strategies discussed by researchers in the literature scored the highest against Layer 7 of the OSI Stack with Layers 4 and 6 scoring the lowest cited works. There were very few studies regarding security attack mitigations to application attacks, with smart printers having no cited works and only a few papers for mitigations against attacks to smart TVs, multimedia systems and smartphones. Cloud attack mitigations scored reasonably high, with most studies discussing encryption systems to maintain data security in the cloud for smart gas and electric meters.

Device attack mitigations were the third most cited works with various research papers discussing mitigation strategies towards attacks to connected autonomous vehicles (CAVs), drones and routers. Interestingly, an additional mitigation category was identified while reviewing the studies in the literature. This is the Security Information and Event Monitoring (SIEM)-based mitigation strategies. These largely comprise general methods to mitigate security attacks against any device with the consideration that these could be applied to consumer IoT devices. SIEM tools are often used to collect data and to review certain systems to deal with a potential incident. The SIEM-based mitigation strategies scored second as the highest cited works after Network attack mitigations. This is likely due to many researchers dedicating general mitigation strategies in the literature and applying these to consumer IoT devices such as running an IDS on a network to see if an IDS can detect intrusions towards consumer IoT devices on the network.

Of the mitigation strategies to security attacks to consumer IoT devices, most published works were done in an Experimental (Lab-based) or Experimental (Simulation) based environment. Again, this is expected. Adding to the constraints mentioned in the previous section, mitigations are harder to implement on devices that are already of public use as it carries security risks and can only be done by researchers that are part of the company producing the devices. Ethical considerations are likely that researchers were unable to attempt to implement these solutions on devices in the consumer domain due to the possibility that the researcher might unintentionally damage a consumer's device and/or collect data about the owner of the participant.

**Crime threats discussed in the literature**

In the results section, crimes were discussed in terms of the mechanisms used, the crimes these mechanisms could facilitate, and the evolving CaaS ecosystem. The latter is important because it lowers the potential barrier to entry, making it easier for a much larger population of offenders to commit the types of crimes discussed. As an example, a Domestic Abuse perpetrator who has no technical background could employ the services of a technically skilled individual to spy on a victim though a device such as a Smart Camera and then relay this feed to the unskilled perpetrator in exchange for a fee. This would mean that there would be a technically skilled individual in the middle of the crime between the offender and the victim.

**Limitations**

The main limitations that could have presented from the systematic review of the literature concerning consumer IoT devices could be that articles were missed due to search term limitations. As the search terms were decided before and make use of the Boolean "AND" and "OR" operators this could potentially limit the results identified of the selected literature. Also, only studies written in English were considered, meaning that relevant studies published in other languages will have been omitted. Future research might seek to include such studies.

5. **CONCLUSIONS**

The conclusions drawn from the work conducted in this study can be divided into four groups. These are, the problem, what was done in this study, the findings and future work for researchers to undertake. These are discussed in the following sections.

**The Problem**

Consumer IoT devices are increasing in popularity, with the ever-increasing expansion of consumer IoT device application types also increasing. However, many studies do not identify the research works regarding the cyber security attack space of these devices. One study was conducted in 2017 [4]. However, the application and device popularity of these devices have expanded in more recent years. Additionally, the cyber security attack space has increased and evolved to take advantage of the new devices and their increased application scenarios. As such many more crimes would need to be considered, in response to this. Additionally, mitigation would also need to be considered as an appropriate response to these new attacks that could facilitate cybercrime.

**What was done?**

With the problems described above, a more recent systematic review was conducted towards crimes that could be facilitated by security vulnerabilities to consumer IoT devices. As such the systematic review incorporates findings in four key areas. These are attacks/security vulnerabilities, mitigation strategies, mechanisms to facilitate cybercrimes and potential cybercrimes. The attacks were modelled using the Open Systems Interconnection (OSI) model, to indicate the layer of the consumer IoT device targeted by these attacks. Additionally, the attacks were modelled against the Hierarchy of Evidence (HoE) to outline how the studies identified in the literature, applied their methodologies when completing the study (i.e. did researchers operate an attack study in a computer lab-based or computer simulated environment, etc.). The OSI model and HoE models were also used when assessing studies related to the countermeasures to attacks for the studies in the literature.

**Main Findings**

The main findings from the systematic review conducted were as follows. Research works discussing cyber security attacks to general consumer IoT devices scored higher over device-specific cyber-attacks. Layer 3 of the OSI model was the most studied cyber-attack category for general cyber-attacks to consumer IoT devices. The least studied layer of the OSI model was Layer 2. Layer 1 was the second most targeted Layer by researchers, likely because researchers like to test cyber-attacks to specific device components (i.e. Wi-Fi, WSN, etc.). The HoE categories showed that research works that discussed Experimental (Lab-based) attacks followed very closely by Experimental (Simulation) were the most discussed research works. This is to be expected as there are likely ethical implications and practicality implications. The least studied research works were Real-world based scenarios.

Research works regarding countermeasures towards cyber-attacks that could facilitate crime through consumer IoT devices were also discussed significantly by researchers. Again, countermeasures towards cyber-attacks targeted towards general consumer IoT devices were discussed the most. Again, the most studied Layer of the OSI model for countermeasures was Layer 3. The least studied Layer of the OSI model was Layer 5. The mitigations saw an extra category of mitigation strategies, in addition, to Device, Network, Cloud and Application attack mitigations. This was Security Information and Event Monitoring (SIEM)-based countermeasures. This scored the second highest category of mitigations discussed in the literature, when applied to consumer IoT devices. Again, Experimental (Lab-based), followed by Experimental (Simulation) HoE categorizations scored the highest research works. Again, this is to be expected due to the reasons outlined above. Real-world again score the lowest research works.

Finally, many cybercrime mechanisms and cybercrimes were identified in the literature. Research works again focussed largely on the Experimental (Lab-based) and Experimental (Simulation) HoE scenarios. Again, this is to be expected as crimes, while they are likely happening already, the data is not available in a real-world environment. So many researchers make assumptions based on the cyber-attacks they test in experimental settings as part of their research.

**Future Work**

Many of the examples found from the review of the literature identified research works that did not implement the methodologies in the real-world domain. As discussed, this is due to several reasons, however understanding to which extent the findings were applicable in the real world is an important future direction. Moreover, this analysis may lead to guidelines and relevant opportunities for researchers to understand the pitfalls of lab based or simulated environments and increase, if needed, the quality of future research.

Another aspect that was noticeable in our review is that often researchers focus on attacks against specific devices or general attacks that may work even outside the IoT domain; the same strategy is also applied to defences and mitigations. It would be interesting in the future to evaluate groups of devices made for the same purpose (e.g. a large set of smart doorbells made by different manufacturers). This would allow to understand which devices may be subject to different type of threats and which capabilities would be needed.

**REFERENCES**

[1] Transforma Insights. (2025). Number of Internet of Things (IoT) connected devices worldwide from 2019 to 2034, by vertical (in millions) [Graph]. Retrieved 9th July 2025, from https://www.statista.com/statistics/1194682/iot-connected-devices-vertically/
[2] Cohen, L. E., & Felson, M. (1979). Social change and crime rate trends: a routine activity approach. American Sociologial Review, 44 (4), 588-608.
[3] Gemalto. (2017). State of IoT Security Research Report from https://www6.thalesgroup.com/state-of-iot-security-2017-press-release
[4] Blythe, J. M., & Johnson, S. D. A systematic review of crime facilitated by the consumer Internet of Things. *Security Journal, 34*(1), 97-125. (2021)
[5] Buil-Gil, D., Kemp, S., Kuenzel, S., et al. The digital harms of smart home devices: A systematic literature review. *Computers in Human Behavior, 145*, 107770. (2023)
[6] Soni, V. D. Security issues in using iot enabled devices and their Impact. *Int. Eng. J. Res. Dev, 4*(2), 7. (2019)
[7] Marton, A. (2023). IoT malware attacks up by 37% in the first half of 2023., from https://iotac.eu/iot-malware-attacks-up-by-37-in-the-first-half-of-2023/
[8] Agrawal, S., Oza, P., Kakkar, R., et al. Analysis and recommendation system-based on PRISMA checklist to write systematic review. *Assessing Writing, 61*, 100866. (2024)

[9] Nishikawa-Pacher, A. Research Questions with PICO: A Universal Mnemonic. *Publications, 10*(3), 21. (2022)

[10] Hallgren, K. A. Computing inter-rater reliability for observational data: an overview and tutorial. *Tutorials in quantitative methods for psychology, 8*(1), 23. (2012)

[11] Higgins, J. P., Thomas, J., Chandler, J., et al. (2019). *Cochrane handbook for systematic reviews of interventions*: John Wiley & Sons.

[12] Javadi, M., & Zarea, K. Understanding thematic analysis and its pitfall. *Journal of client care, 1*(1), 33-39. (2016)

[13] Allifah, N. M., & Zualkernan, I. A. Ranking Security of IoT-Based Smart Home Consumer Devices. *IEEE Access, 10*, 18352-18369. (2022)

[14] Alqassem, I. (2014). *Privacy and security requirements framework for the internet of things (IoT)*. Paper presented at the Companion Proceedings of the 36th International Conference on Software Engineering, Hyderabad, India. https://doi.org/10.1145/2591062.2591201

[15] Giri, A., Dutta, S., Neogy, S., Dahal, K., & Pervez, Z. (2017). *Internet of things (IoT): a survey on architecture, enabling technologies, applications and challenges*. Paper presented at the Proceedings of the 1st International Conference on Internet of Things and Machine Learning, Liverpool, United Kingdom. https://doi.org/10.1145/3109761.3109768

[16] Aggarwal, R., & Das, M. L. (2012). *RFID security in the context of "internet of things"*. Paper presented at the Proceedings of the First International Conference on Security of Internet of Things, Kollam, India. https://doi.org/10.1145/2490428.2490435

[17] R. Al, M., M. Al, A., & S. El, K. (2020, 9-10 Sept. 2020). *IoT: Security Challenges and Issues of Smart Homes/Cities.* Paper presented at the 2020 International Conference on Computing and Information Technology (ICCIT-1441).

[18] Chaudhary, R., & Ragiri, P. R. (2016). *Implementation and Analysis of Blackhole Attack in AODV Routing Protocol*. Paper presented at the Proceedings of the Second International Conference on Information and Communication Technology for Competitive Strategies, Udaipur, India. https://doi.org/10.1145/2905055.2905172

[19] Kaur, K., & Raj, G. (2012). *Comparative analysis of Black Hole attack over Cloud Network using AODV and DSDV*. Paper presented at the Proceedings of the Second International Conference on Computational Science, Engineering and Information Technology, Coimbatore UNK, India. https://doi.org/10.1145/2393216.2393334

[20] Tatang, D., Suurland, T., & Holz, T. (2020). *Study of DNS Rebinding Attacks on Smart Home Devices*. Paper presented at the COMPUTER SECURITY, ESORICS 2019.

[21] Sudharsan, B., Sundaram, D., Patel, P., Breslin, J. G., & Ali, M. I. (2021, 22-26 March 2021). *Edge2Guard: Botnet Attacks Detecting Offline Models for Resource-Constrained IoT Devices.* Paper presented at the 2021 IEEE International Conference on Pervasive Computing and Communications Workshops and other Affiliated Events (PerCom Workshops).

[22] Svaigen, A. R., Boukerche, A., Ruiz, L. B., & Loureiro, A. A. F. (2022). *Is the Remote ID a Threat to the Drone's Location Privacy on the Internet of Drones?* Paper presented at the Proceedings of the 20th ACM International Symposium on Mobility Management and Wireless Access, Montreal, Quebec, Canada. https://doi.org/10.1145/3551660.3560914

[23] Laaroussi, Z., Morabito, R., & Taleb, T. (2018, 29-31 Oct. 2018). *Service Provisioning in Vehicular Networks Through Edge and Cloud: An Empirical Analysis.* Paper presented at the 2018 IEEE Conference on Standards for Communications and Networking (CSCN).

[24] Zhang, T., Antunes, H., & Aggarwal, S. Defending Connected Vehicles Against Malware: Challenges and a Solution Framework. *IEEE Internet of Things Journal, 1*(1), 10-21. (2014)

[25] Poornachandran, P., Sreeram, R., Krishnan, M. R., Pal, S., Sankar, A. U. P., & Ashok, A. (2015, 21-23 Dec. 2015). *Internet of Vulnerable Things (IoVT): Detecting Vulnerable SOHO Routers.* Paper presented at the 2015 International Conference on Information Technology (ICIT).

[26] Cheng, K., Li, Q., Wang, L., et al. (2018, 25-28 June 2018). *DTaint: Detecting the Taint-Style Vulnerability in Embedded Device Firmware.* Paper presented at the 2018 48th Annual IEEE/IFIP International Conference on Dependable Systems and Networks (DSN).

[27] Santani, A., Gangaramani, M., Chopra, B., Choudhary, P., & Samdani, K. (2021, 8-10 July 2021). *An Overview of Architecture and Security Issues of a Smart TV.* Paper presented at the 2021 6th International Conference on Communication and Electronics Systems (ICCES).

[28] Overstreet, D., Wimmer, H., & Haddad, R. J. (2019, 11-14 April 2019). *Penetration Testing of the Amazon Echo Digital Voice Assistant Using a Denial-of-Service Attack.* Paper presented at the 2019 SoutheastCon.

[29] Feng, H., & Shin, K. G. (2016). *Understanding and defending the binder attack surface in Android*. Paper presented at the Proceedings of the 32nd Annual Conference on Computer Security Applications, Los Angeles, California, USA. https://doi.org/10.1145/2991079.2991120

[30] Mäki, P., Rauti, S., Hosseinzadeh, S., Koivunen, L., & Leppänen, V. (2016, 6-9 Dec. 2016). *Interface Diversification in IoT Operating Systems.* Paper presented at the 2016 IEEE/ACM 9th International Conference on Utility and Cloud Computing (UCC).

[31] Liu, Y., Hu, S., & Ho, T. Y. (2014, 2-6 Nov. 2014). *Vulnerability assessment and defense technology for smart home cybersecurity considering pricing cyberattacks.* Paper presented at the 2014 IEEE/ACM International Conference on Computer-Aided Design (ICCAD).

[32] Al-Syouf, R., Al-Duwairi, B., & Shatnawi, A. S. (2021). *Towards a Secure Web-Based Smart Homes*. Paper presented at the 2021 12TH INTERNATIONAL CONFERENCE ON INFORMATION AND COMMUNICATION SYSTEMS (ICICS).

[33] Roshani, M., & Nobakht, M. (2022). *HybridDAD: Detecting DDoS Flooding Attack using Machine Learning with Programmable Switches*. Paper presented at the Proceedings of the 17th International Conference on Availability, Reliability and Security, Vienna, Austria. https://doi.org/10.1145/3538969.3538991

[34] Man, K., Zhou, X. a., & Qian, Z. (2021). *DNS Cache Poisoning Attack: Resurrections with Side Channels*. Paper presented at the Proceedings of the 2021 ACM SIGSAC Conference on Computer and Communications Security, Virtual Event, Republic of Korea. https://doi.org/10.1145/3460120.3486219

[35] Zhao, Y., Wang, X., Jiang, Y., Mei, Y., Singh, A. K., & Mak, T. (2018, 4-7 Sept. 2018). *On a New Hardware Trojan Attack on Power Budgeting of Many Core

*Systems*. Paper presented at the 2018 31st IEEE International System-on-Chip Conference (SOCC).

[36] Zahra, S. R., & Chishti, M. A. (2019, 10-11 Jan. 2019). *RansomWare and Internet of Things: A New Security Nightmare.* Paper presented at the 2019 9th International Conference on Cloud Computing, Data Science & Engineering (Confluence).

[37] Ghirardello, K., Maple, C., Ng, D., & Kearney, P. (2018, 28-29 March 2018). *Cyber security of smart homes: Development of a reference architecture for attack surface analysis.* Paper presented at the Living in the Internet of Things: Cybersecurity of the IoT - 2018.

[38] Sethi, B. K., Mukherjee, D., Singh, D., Misra, R. K., & Mohanty, S. R. Smart home energy management system under false data injection attack. *International Transactions on Electrical Energy System, 30*(7). (2020)

[39] Cai, M., Wu, Z., & Zhang, J. (2014, 8-10 Nov. 2014). *Research and Prevention of Rogue AP Based MitM in Wireless Network.* Paper presented at the 2014 Ninth International Conference on P2P, Parallel, Grid, Cloud and Internet Computing.

[40] Setiawan, F. B., & Magfirawaty. (2021, 20-21 Oct. 2021). *Securing Data Communication Through MQTT Protocol with AES-256 Encryption Algorithm CBC Mode on ESP32-Based Smart Homes.* Paper presented at the 2021 International Conference on Computer System, Information Technology, and Electrical Engineering (COSITE).

[41] Thombare, B. M., & Soni, D. R. (2022). *Prevention of SQL Injection Attack by Using Black Box Testing*. Paper presented at the Proceedings of the 23rd International Conference on Distributed Computing and Networking, Delhi, AA, India. https://doi.org/10.1145/3491003.3493233

[42] Yu, S., Das, A. K., & Park, Y. Comments on "ALAM: Anonymous Lightweight Authentication Mechanism for SDN Enabled Smart Homes". *IEEE Access, 9*, 49154-49159. (2021)

[43] Yu, S., Jho, N., & Park, Y. Lightweight Three-Factor-Based Privacy- Preserving Authentication Scheme for IoT-Enabled Smart Homes. *IEEE Access, 9*, 126186-126197. (2021)

[44] Coppolino, L., Alessandro, V. D., Antonio, S. D., Levy, L., & Romano, L. (2015, 21-23 Oct. 2015). *My Smart Home is Under Attack.* Paper presented at the 2015 IEEE 18th International Conference on Computational Science and Engineering.

[45] Durvaux, F., & Durvaux, M. SCA-Pitaya: A Practical and Affordable Side-Channel Attack Setup for Power Leakage--Based Evaluations. *Digital Threats, 1*(1), Article 3. (2020)

[46] Abrishamchi, M. A. N., Abdullah, A. H., Cheok, A. D., Bielawski, K. S., & Ieee. (2017). *Side Channel Attacks on Smart Home Systems: A Short Overview*. Paper presented at the IECON 2017 - 43RD ANNUAL CONFERENCE OF THE IEEE INDUSTRIAL ELECTRONICS SOCIETY.

[47] Ding, R., Zhang, Z., Zhang, X., Gongye, C., Fei, Y., & Ding, A. A. (2022). *A cross-platform cache timing attack framework via deep learning*. Paper presented at the Proceedings of the 2022 Conference & Exhibition on Design, Automation & Test in Europe, Antwerp, Belgium.

[48] Huraj, L., Simon, M., & Horak, T. Resistance of IoT Sensors against DDoS Attack in Smart Home Environment. *SENSORS, 20*(18). (2020)

[49] Lejla, B., Nele, M., Markus, M., et al. (2022). Advanced Attacks and Protection Mechanisms in IoT Devices and Networks *Security and Privacy in the Internet of Things: Architectures, Techniques, and Applications* (pp. 1-30): IEEE.

[50] Santos, B. V. D., Vergütz, A., Macedo, R. T., & Nogueira, M. (2022, 30 Nov.-2 Dec. 2022). *A Dynamic Method to Protect User Privacy Against Traffic-based Attacks on Smart Home.* Paper presented at the 2022 IEEE Latin-American Conference on Communications (LATINCOM).

[51] Bellman, C., & Oorschot, P. C. v. (2019, 26-28 Aug. 2019). *Analysis, Implications, and Challenges of an Evolving Consumer IoT Security Landscape.* Paper presented at the 2019 17th International Conference on Privacy, Security and Trust (PST).

[52] Rajendran, G., Nivash, R. S. R., Parthy, P. P., & Balamurugan, S. (2019, 1-3 Oct. 2019). *Modern security threats in the Internet of Things (IoT): Attacks and Countermeasures.* Paper presented at the 2019 International Carnahan Conference on Security Technology (ICCST).

[53] Nobakht, M., Mahmoudi, H., & Rahimzadeh, O. (2022). *A Distributed Security Approach against ARP Cache Poisoning Attack*. Paper presented at the Proceedings of the 1st Workshop on Cybersecurity and Social Sciences, Nagasaki, Japan. https://doi.org/10.1145/3494108.3522765

[54] Anthi, E., Williams, L., Słowińska, M., Theodorakopoulos, G., & Burnap, P. A Supervised Intrusion Detection System for Smart Home IoT Devices. *IEEE Internet of Things Journal, 6*(5), 9042-9053. (2019)

[55] Patil, R., & Tahiliani, M. P. (2014, 19-20 Aug. 2014). *Detecting packet modification attack by misbehaving router.* Paper presented at the 2014 First International Conference on Networks & Soft Computing (ICNSC2014).

[56] Ou, C. W., Hsu, F. H., & Lai, C. M. (2019, 5-8 Aug. 2019). *Keep Rogue IoT Away: IoT Detector Based on Diversified TLS Negotiation.* Paper presented at the 2019 IEEE Intl Conf on Dependable, Autonomic and Secure Computing, Intl Conf on Pervasive Intelligence and Computing, Intl Conf on Cloud and Big Data Computing, Intl Conf on Cyber Science and Technology Congress (DASC/PiCom/CBDCom/CyberSciTech).

[57] Zolotykh, M. (2020, 17-19 Nov. 2020). *Research of Behavior of the Search Engine 'Shodan.io'.* Paper presented at the 2020 Global Smart Industry Conference (GloSIC).

[58] Lan, L., Xinping, H., Hanchun, C., & Jiancheng, D. (2016, 21-23 July 2016). *Non-repudiation Scheme of Smart Home Based on Cloud Service.* Paper presented at the 2016 Sixth International Conference on Instrumentation & Measurement, Computer, Communication and Control (IMCCC).

[59] Almutairi, O., & Almarhabi, K. Investigation of Smart Home Security and Privacy: Consumer Perception in Saudi Arabia. *International Journal of Advanced Computer Science and Applications, 12*(4). (2021)

[60] Weiss, N., Renner, S., Mottok, J., & Matoušek, V. (2020). *Transport Layer Scanning for Attack Surface Detection in Vehicular Networks*. Paper presented at the Proceedings of the 4th ACM Computer Science in Cars Symposium, Feldkirchen, Germany. https://doi.org/10.1145/3385958.3430476

[61] Morel, A. E., Ufuktepe, D. K., Ignatowicz, R., et al. (2020, 13-15 Oct. 2020). *Enhancing Network-edge Connectivity and Computation Security in Drone Video Analytics.* Paper presented at the 2020 IEEE Applied Imagery Pattern Recognition Workshop (AIPR).

[62] Visoottiviseth, V., Jutadhammakorn, P., Pongchanchai, N., & Kosolyudhthasarn, P. (2018, 11-13 July 2018). *Firmaster: Analysis Tool for Home Router Firmware.* Paper presented at the 2018 15th International Joint Conference on Computer Science and Software Engineering (JCSSE).

[63] Mao, J., Zhu, S., Dai, X., Lin, Q., & Liu, J. Watchdog: Detecting Ultrasonic-Based Inaudible Voice Attacks to Smart Home Systems. *IEEE Internet of Things Journal, 7*(9), 8025-8035. (2020)

[64] Privitera, D., & Shahriar, H. (2018, 5-7 June 2018). *Design and Development of Smart TV Protector.* Paper presented at the 2018 National Cyber Summit (NCS).

[65] Kantola, D., Chin, E., He, W., & Wagner, D. (2012). *Reducing attack surfaces for intra-application communication in android*. Paper presented at the Proceedings of the second ACM workshop on Security and privacy in smartphones and mobile devices, Raleigh, North Carolina, USA. https://doi.org/10.1145/2381934.2381948

[66] Zhou, Y., Liu, Y., & Hu, S. Smart Home Cyberattack Detection Framework for Sponsor Incentive Attacks. *IEEE Transactions on Smart Grid, 10*(2), 1916-1927. (2019)

[67] Azumah, S. W., Elsayed, N., Adewopo, V., Zaghloul, Z. S., & Li, C. (2021, 14 June-31 July 2021). *A Deep LSTM based Approach for Intrusion Detection IoT Devices Network in Smart Home.* Paper presented at the 2021 IEEE 7th World Forum on Internet of Things (WF-IoT).

[68] Goel, A., Patel, J., & Patel, C. (2022, 3-5 Oct. 2022). *BLS based authentication and token-based authorization for Smart Home.* Paper presented at the 2022 13th International Conference on Computing Communication and Networking Technologies (ICCCNT).

[69] Yan, Y., Tan, J., & Tang, Q. (2022). *Design and Analysis of Internet of Things Application Cryptosystem Based on Blockchain Technology*. Paper presented at the Proceedings of the 3rd Asia-Pacific Conference on Image Processing, Electronics and Computers, Dalian, China. https://doi.org/10.1145/3544109.3544348

[70] Ameer, S., Benson, J., & sandhu, R. Hybrid Approaches (ABAC and RBAC) Toward Secure Access Control in Smart Home IoT. *IEEE Transactions on Dependable and Secure Computing*, 1-18. (2022)

[71] Li, R., Li, Q., Zhou, J., & Jiang, Y. ADRIoT: An Edge-assisted Anomaly Detection Framework against IoT-based Network Attacks. *IEEE Internet of Things Journal*, 1-1. (2021)

[72] Mustapha, H., & Alghamdi, A. M. (2018). *DDoS attacks on the internet of things and their prevention methods*. Paper presented at the Proceedings of the 2nd International Conference on Future Networks and Distributed Systems, Amman, Jordan. https://doi.org/10.1145/3231053.3231057

[73] Loi, F., Sivanathan, A., Gharakheili, H. H., Radford, A., & Sivaraman, V. (2017). *Systematically Evaluating Security and Privacy for Consumer IoT Devices*. Paper presented at the Proceedings of the 2017 Workshop on Internet of Things Security and Privacy, Dallas, Texas, USA. https://doi.org/10.1145/3139937.3139938

[74] Jonsdottir, G., Wood, D., & Doshi, R. (2017, 3-5 Nov. 2017). *IoT network monitor.* Paper presented at the 2017 IEEE MIT Undergraduate Research Technology Conference (URTC).

[75] Ryoo, J., Tjoa, S., & Ryoo, H. (2018, 26-27 July 2018). *An IoT Risk Analysis Approach for Smart Homes (Work-in-Progress).* Paper presented at the 2018 International Conference on Software Security and Assurance (ICSSA).

[76] Gupta, N., Naik, V., & Sengupta, S. (2017, 4-8 Jan. 2017). *A firewall for Internet of Things.* Paper presented at the 2017 9th International Conference on Communication Systems and Networks (COMSNETS).

[77] Memos, V. A., & Psannis, K. E. (2020, 9-11 Oct. 2020). *AI-Powered Honeypots for Enhanced IoT Botnet Detection.* Paper presented at the 2020 3rd World Symposium on Communication Engineering (WSCE).

[78] Mendonça, G., Santos, G. H. A., Silva, E. d. S. e., Leão, R. M. M., Menasché, D. S., & Towsley, D. (2019, 9-12 Dec. 2019). *An Extremely Lightweight Approach for DDoS Detection at Home Gateways.* Paper presented at the 2019 IEEE International Conference on Big Data (Big Data).

[79] Auliar, R. B., & Bekaroo, G. (2021, 7-8 Oct. 2021). *Security in IoT-based Smart Homes: A Taxonomy Study of Detection Methods of Mirai Malware and Countermeasures.* Paper presented at the 2021 International Conference on Electrical, Computer, Communications and Mechatronics Engineering (ICECCME).

[80] Alsabilah, N., & Rawat, D. B. (2023). An Adaptive Flow-based NIDS for Smart Home Networks Against Malware Behavior Using XGBoost combined with Rough Set Theory. Piscataway: The Institute of Electrical and Electronics Engineers, Inc. (IEEE).

[81] Popoola, S. I., Adebisi, B., Hammoudeh, M., Gacanin, H., & Gui, G. Stacked recurrent neural network for botnet detection in smart homes. *COMPUTERS & ELECTRICAL ENGINEERING, 92*. (2021)

[82] Malkawe, R., Qasaimeh, M., Ghanim, F., & Ababneh, M. (2019). *Toward an early assessment for Ransomware attack vulnerabilities*. Paper presented at the Proceedings of the Second International Conference on Data Science, E-Learning and Information Systems, Dubai, United Arab Emirates. https://doi.org/10.1145/3368691.3368734

[83] Anand, P. M., Charan, P. V. S., & Shukla, S. K. HiPeR - Early Detection of a Ransomware Attack using Hardware Performance Counters. *Digital Threats, 4*(3), Article 43. (2023)

[84] AbdulsalamYa'u, G., Job, G. K., Waziri, S. M., Jaafar, B., SabonGari, N. A., & Yakubu, I. Z. (2019, 13-14 Dec. 2019). *Deep Learning for Detecting Ransomware in Edge Computing Devices Based On Autoencoder Classifier.* Paper presented at the 2019 4th International Conference on Electrical, Electronics, Communication, Computer Technologies and Optimization Techniques (ICEECCOT).

[85] Imdad, U., Habib, M. A., Ahmad, M., Mahmood, N., & Ashraf, R. (2017). *Auto Configuration Based Enhanced and Secure Domain Naming Service for IPV-6 Internet of Things*. Paper presented at the Proceedings of the International Conference on Future Networks and Distributed Systems, Cambridge, United Kingdom. https://doi.org/10.1145/3102304.3109818

[86] Aizuddin, A. A., Atan, M., Norulazmi, M., Noor, M. M., Akimi, S., & Abidin, Z. (2017). *DNS amplification attack detection and mitigation via sFlow with security-centric SDN*. Paper presented at the Proceedings of the 11th International Conference on Ubiquitous Information Management and Communication, Beppu, Japan. https://doi.org/10.1145/3022227.3022230

[87] Li, Y., Wang, Y. Z., & Zhang, Y. (2018). *SecHome: A Secure Large-Scale Smart Home System Using Hierarchical Identity Based Encryption*. Paper presented at the INFORMATION AND COMMUNICATIONS SECURITY, ICICS 2017.

[88] Alsmady, A., Alazzam, H., & Al-Shorman, A. (2019). *Black hole attack prevention in MANET using enhanced AODV protocol*. Paper presented at the Proceedings of the Second International Conference on Data Science, E-Learning and Information Systems, Dubai, United Arab Emirates. https://doi.org/10.1145/3368691.3368732

[89] Rai, A., Patel, R., Kapoor, R. K., & Karaulia, D. S. (2014). *Enhancement in Security of AODV Protocol against Black-hole Attack in MANET*. Paper presented at the Proceedings of the 2014 International Conference on Information and Communication Technology for Competitive Strategies, Udaipur, Rajasthan, India. https://doi.org/10.1145/2677855.2677946

[90] Malik, D., & Rizvi, M. A. (2014). *Prevention of Node Isolation Attack on OLSR by DFOLSR*. Paper presented at the Proceedings of the 2014 International Conference on Information and Communication Technology for Competitive Strategies, Udaipur, Rajasthan, India. https://doi.org/10.1145/2677855.2677885

[91] Chen, K., Gu, L., & Sun, J. (2020). *A DoS Attack Detection Method Based on Multi-source Data Fusion*. Paper presented at the Proceedings of the 4th International Conference on Computer Science and Application Engineering, Sanya, China. https://doi.org/10.1145/3424978.3425098

[92] Gattu, N., Khan, M. N. I., De, A., & Ghosh, S. (2020). *Power side channel attack analysis and detection*. Paper presented at the Proceedings of the 39th International Conference on Computer-Aided Design, Virtual Event, USA. https://doi.org/10.1145/3400302.3415692

[93] Cavalcante, J., Trajano, A. F. R., Leite, L., et al. (2019, 8-11 Sept. 2019). *Securing IPv6 Wireless Networks Against Malicious Router Advertisements.* Paper presented at the 2019 IEEE 30th Annual International Symposium on Personal, Indoor and Mobile Radio Communications (PIMRC).

[94] Li, J., Zhang, N., Ni, J., Chen, J., & Du, R. Secure and Lightweight Authentication With Key Agreement for Smart Wearable Systems. *IEEE Internet of Things Journal, 7*(8), 7334-7344. (2020)

[95] Naoui, S., Elhdhili, M. H., & Saidane, L. A. (2019, 15-18 April 2019). *Novel Smart Home Authentication Protocol LRP-SHAP.* Paper presented at the 2019 IEEE Wireless Communications and Networking Conference (WCNC).

[96] Nyangaresi, V. O., & Ogundoyin, S. O. (2021, 5-8 Oct. 2021). *Certificate Based Authentication Scheme for Smart Homes.* Paper presented at the 2021 3rd Global Power, Energy and Communication Conference (GPECOM).

[97] Rana Muhammad Abdul, H.-u.-r., Liaqat, M., Azana Hafizah Mohd, A., et al. LR-AKAP: A Lightweight and Robust Security Protocol for Smart Home Environments. *SENSORS, 22*(18), 6902. (2022)

[98] Shen, T., & Ma, M. (2016, 22-25 Nov. 2016). *Security enhancements on Home Area Networks in smart grids.* Paper presented at the 2016 IEEE Region 10 Conference (TENCON).

[99] Bang, A. O., & Rao, U. P. A novel decentralized security architecture against sybil attack in RPL-based IoT networks: a focus on smart home use case. *The Journal of Supercomputing, 77*(12), 13703-13738. (2021)

[100] Abdelhamid, A., Elsayed, M. S., Aslan, H. K., & Azer, M. A. (2024). *Anomaly-Based Intrusion Detection for Blackhole Attack Mitigation*. Paper presented at

the Proceedings of the 19th International Conference on Availability, Reliability and Security, Vienna, Austria. https://doi.org/10.1145/3664476.3670941

[101] Daoud, L., & Rafla, N. (2019). *Detection and prevention protocol for black hole attack in network-on-chip*. Paper presented at the Proceedings of the 13th IEEE/ACM International Symposium on Networks-on-Chip, New York, New York. https://doi.org/10.1145/3313231.3352374

[102] Yerneni, R., & Sarje, A. K. (2012). *Enhancing performance of AODV against Black hole Attack*. Paper presented at the Proceedings of the CUBE International Information Technology Conference, Pune, India. https://doi.org/10.1145/2381716.2393516

[103] Jain, S., & Khunteta, A. (2015). *Detection Techniques of Blackhole Attack in Mobile Adhoc Network: A Survey*. Paper presented at the Proceedings of the 2015 International Conference on Advanced Research in Computer Science Engineering & Technology (ICARCSET 2015), Unnao, India. https://doi.org/10.1145/2743065.2743112

[104] Dhagat, R., & Jagtap, P. (2016). *An Approach Used In Wireless Network to Detect Denial of Service Attack*. Paper presented at the Proceedings of the ACM Symposium on Women in Research 2016, Indore, India. https://doi.org/10.1145/2909067.2909069

[105] Fox, P. (2021). *TrustedAP: Using the Ethereum Blockchain to Mitigate the Evil Twin Attack*. Paper presented at the Proceedings of the 52nd ACM Technical Symposium on Computer Science Education, Virtual Event, USA. https://doi.org/10.1145/3408877.3439695

[106] Mbarek, B., Ge, M., & Pitner, T. Trust-Based Authentication for Smart Home Systems. *Wireless Personal Communications, 117*(3), 2157-2172. (2021)

[107] Kasinathan, P., Costamagna, G., Khaleel, H., Pastrone, C., & Spirito, M. A. (2013). *DEMO: An IDS framework for internet of things empowered by 6LoWPAN*. Paper presented at the Proceedings of the 2013 ACM SIGSAC conference on Computer & communications security, Berlin, Germany. https://doi.org/10.1145/2508859.2512494

[108] Benzaïd, C., Boulgheraif, A., Dahmane, F. Z., Al-Nemrat, A., & Zeraoulia, K. (2016). *Intelligent detection of MAC spoofing attack in 802.11 network*. Paper presented at the Proceedings of the 17th International Conference on Distributed Computing and Networking, Singapore, Singapore. https://doi.org/10.1145/2833312.2850446

[109] Tang, Z. Y., Zhao, Y. J., Yang, L., et al. Exploiting Wireless Received Signal Strength Indicators to Detect Evil-Twin Attacks in Smart Homes. *MOBILE INFORMATION SYSTEMS, 2017*. (2017)

[110] Mahapatra, B., Panda, S. K., Turuk, A. K., Sahoo, B., & Patra, S. K. (2019, 16-18 Dec. 2019). *LW-AKA: A Security Protocol for Integrated RFID and IoT Based Smart Home Security System.* Paper presented at the 2019 IEEE International Symposium on Smart Electronic Systems (iSES) (Formerly iNiS).

[111] Yao, Q., Ma, J., Cong, S., Li, X., & Li, J. (2016). *Attack gives me power: DoS-defending constant-time privacy-preserving authentication of low-cost devices such as backscattering RFID tags*. Paper presented at the Proceedings of the 3rd ACM Workshop on Mobile Sensing, Computing and Communication, Paderborn, Germany. https://doi.org/10.1145/2940353.2940361

[112] Xu, Y., Jiang, Y., Hu, C., Chen, H., He, L., & Cao, Y. (2014, 19-23 Oct. 2014). *A balanced security protocol of Wireless Sensor Network for Smart Home.*

Paper presented at the 2014 12th International Conference on Signal Processing (ICSP).

[113] Lu, Y.-F., Kuo, C.-F., Chen, H.-M., Wang, G.-B., & Chou, S.-C. (2018). *A mutual authentication scheme with user anonymity for cyber-physical and internet of things*. Paper presented at the Proceedings of the 2018 Conference on Research in Adaptive and Convergent Systems, Honolulu, Hawaii. https://doi.org/10.1145/3264746.3264762

[114] Arvandy, & Bandung, Y. (2018). *Design of Secure IoT Platform for Smart Home System*. Paper presented at the 2018 5TH INTERNATIONAL CONFERENCE ON INFORMATION TECHNOLOGY, COMPUTER, AND ELECTRICAL ENGINEERING (ICITACEE).

[115] Muhammad, R., & Baseer, S. Authentication and Privacy Challenges for Internet of Things Smart Home Environment. *JOURNAL OF MECHANICS OF CONTINUA AND MATHEMATICAL SCIENCES, 14*(1), 258-275. (2019)

[116] Amadeo, M., Giordano, A., Mastroianni, C., & Molinaro, A. (2019). On the Integration of Information Centric Networking and Fog Computing for Smart Home Services. In F. Cicirelli, A. Guerrieri, C. Mastroianni, G. Spezzano & A. Vinci (Eds.), *INTERNET OF THINGS FOR SMART URBAN ECOSYSTEMS* (pp. 75-93).

[117] Arafin, M. T., Anand, D., & Qu, G. (2017). *A Low-Cost GPS Spoofing Detector Design for Internet of Things (IoT) Applications*. Paper presented at the Proceedings of the on Great Lakes Symposium on VLSI 2017, Banff, Alberta, Canada. https://doi.org/10.1145/3060403.3060455

[118] Alsirhani, A., Khan, M. A., Alomari, A., et al. Securing Low-Power Blockchain-enabled IoT Devices against Energy Depletion Attack. *ACM Trans. Internet Technol., 23*(3), Article 43. (2023)

[119] Mahmud, F., Kim, S., Chawla, H. S., Kim, E. J., Tsai, C.-C., & Muzahid, A. (2023). *Attack of the Knights:Non Uniform Cache Side Channel Attack*. Paper presented at the Proceedings of the 39th Annual Computer Security Applications Conference, Austin, TX, USA. https://doi.org/10.1145/3627106.3627199

[120] Yuan, J., Zhang, M., & Zheng, D. (2025). *Side-channel attack based on CAP model*. Paper presented at the Proceedings of the 2024 7th International Conference on Artificial Intelligence and Pattern Recognition. https://doi.org/10.1145/3703935.3704041

[121] Maleh, Y., Ezzati, A., & Belaissaoui, M. (2016). *DoS Attacks Analysis and Improvement in DTLS Protocol for Internet of Things*. Paper presented at the Proceedings of the International Conference on Big Data and Advanced Wireless Technologies, Blagoevgrad, Bulgaria. https://doi.org/10.1145/3010089.3010139

[122] Berbecaru, D. G., & Lioy, A. (2024). *Threat-TLS: A Tool for Threat Identification in Weak, Malicious, or Suspicious TLS Connections*. Paper presented at the Proceedings of the 19th International Conference on Availability, Reliability and Security, Vienna, Austria. https://doi.org/10.1145/3664476.3670945

[123] Wang, L., Bu, L., & Song, F. (2025). SCAGuard: Detection and Classification of Cache Side-Channel Attacks via Attack Behavior Modeling and Similarity Comparison *Proceedings of the 60th Annual ACM/IEEE Design Automation Conference* (pp. 1–6): IEEE Press.

[124] Shen, F., Zhang, S., Liu, Y., & Yang, Z. (2022, 30 May-3 June 2022). *A Survey of Traffic Obfuscation Technology for Smart Home.* Paper presented at the

2022 International Wireless Communications and Mobile Computing (IWCMC).
[125] Bhade, P., Paturel, J., Sentieys, O., & Sinha, S. Lightweight Hardware-Based Cache Side-Channel Attack Detection for Edge Devices (Edge-CaSCADe). *ACM Trans. Embed. Comput. Syst., 23*(4), Article 56. (2024)
[126] Chang, Y., Yan, Y., Zhu, C., & Liu, Y. A High-performance Masking Design Approach for Saber against High-order Side-channel Attack. *ACM Trans. Des. Autom. Electron. Syst., 28*(6), Article 91. (2023)
[127] Chen, Y., Hajiabadi, A., Poussier, R., et al. PARADISE: Criticality-Aware Instruction Reordering for Power Attack Resistance. *ACM Trans. Archit. Code Optim., 22*(1), Article 14. (2025)
[128] Bhattacharya, S., Manousakas, D., Ramos, A. G. C. P., Venieris, S. I., Lane, N. D., & Mascolo, C. Countering Acoustic Adversarial Attacks in Microphone-equipped Smart Home Devices. *Proc. ACM Interact. Mob. Wearable Ubiquitous Technol., 4*(2), Article 73. (2020)
[129] Nyangaresi, V. O. (2021, 13-15 Sept. 2021). *Lightweight Key Agreement and Authentication Protocol for Smart Homes.* Paper presented at the 2021 IEEE AFRICON.
[130] Pan, Y., Xu, Z., Li, M., & Lazos, L. (2021). *Man-in-the-Middle Attack Resistant Secret Key Generation via Channel Randomization*. Paper presented at the Proceedings of the Twenty-second International Symposium on Theory, Algorithmic Foundations, and Protocol Design for Mobile Networks and Mobile Computing, Shanghai, China. https://doi.org/10.1145/3466772.3467052
[131] Ahmed, A. S. A. M. S., Hassan, R., & Othman, N. E. IPv6 Neighbor Discovery Protocol Specifications, Threats and Countermeasures: A Survey. *IEEE Access, 5*, 18187-18210. (2017)
[132] Al-Shaboti, M., Welch, I., Chen, A., & Mahmood, M. A. (2018). *Towards Secure Smart Home IoT: Manufacturer and User Network Access Control Framework*. Paper presented at the PROCEEDINGS 2018 IEEE 32ND INTERNATIONAL CONFERENCE ON ADVANCED INFORMATION NETWORKING AND APPLICATIONS (AINA).
[133] Rajora, C. S., & Sharma, A. (2022, 16-17 Dec. 2022). *IoT Based Smart Home with Cutting-Edge Technology for IDS/IPS.* Paper presented at the 2022 Second International Conference on Advanced Technologies in Intelligent Control, Environment, Computing & Communication Engineering (ICATIECE).
[134] Loukas, G., Vuong, T., Heartfield, R., Sakellari, G., Yoon, Y., & Gan, D. Cloud-Based Cyber-Physical Intrusion Detection for Vehicles Using Deep Learning. *IEEE Access, 6*, 3491-3508. (2018)
[135] Trabelsi, Z. (2012). *Switch's CAM table poisoning attack: hands-on lab exercises for network security education*. Paper presented at the Proceedings of the Fourteenth Australasian Computing Education Conference - Volume 123, Melbourne, Australia.
[136] Salami, S. A., Baek, J., Salah, K., & Damiani, E. (2016, 31 Aug.-2 Sept. 2016). *Lightweight Encryption for Smart Home.* Paper presented at the 2016 11th International Conference on Availability, Reliability and Security (ARES).
[137] Naoui, S., Elhdhili, M. E., & Saidane, L. A. (2017, 17-21 July 2017). *Lightweight Enhanced Collaborative Key Management Scheme for Smart Home Application.* Paper presented at the 2017 International Conference on High Performance Computing & Simulation (HPCS).

[138] Kvamme, S. M., Gudmundsen, E., Oyetoyan, T. D., & Cruzes, D. S. (2023). *Data Protection Fortification: An Agile Approach for Threat Analysis of IoT Data*. Paper presented at the Proceedings of the 12th International Conference on the Internet of Things, Delft, Netherlands. https://doi.org/10.1145/3567445.3569164

[139] Mundt, M., & Baier, H. Threat-Based Simulation of Data Exfiltration Toward Mitigating Multiple Ransomware Extortions. *Digital Threats, 4*(4), Article 54. (2023)

[140] Shakhov, V., Nam, S., & Choo, H. (2013). *Flooding attack in energy harvesting wireless sensor networks*. Paper presented at the Proceedings of the 7th International Conference on Ubiquitous Information Management and Communication, Kota Kinabalu, Malaysia. https://doi.org/10.1145/2448556.2448605

[141] Wang, J., Fei, J., Zhu, Y., Wang, X., & Lin, X. (2025). *Link Flooding Attack Mitigation Method Based on SDN Service Priority*. Paper presented at the Proceedings of the 2025 4th International Conference on Intelligent Systems, Communications and Computer Networks. https://doi.org/10.1145/3732945.3732980

[142] Hakim, M. A., Aksu, H., Uluagac, A. S., & Akkaya, K. (2018, 17-19 Nov. 2018). *U-PoT: A Honeypot Framework for UPnP-Based IoT Devices*. Paper presented at the 2018 IEEE 37th International Performance Computing and Communications Conference (IPCCC).

[143] Sousa, B. F. L. M., Abdelouahab, Z., Lopes, D. C. P., Soeiro, N. C., & Ribeiro, W. F. (2017). *An intrusion detection system for denial of service attack detection in internet of things*. Paper presented at the Proceedings of the Second International Conference on Internet of things, Data and Cloud Computing, Cambridge, United Kingdom. https://doi.org/10.1145/3018896.3018962

[144] Procopiou, A., Komninos, N., & Douligeris, C. ForChaos: Real Time Application DDoS Detection Using Forecasting and Chaos Theory in Smart Home IoT Network. *Wireless Communications & Mobile Computing (Online), 2019*, 14. (2019)

[145] Kashyap, H. J., & Bhattacharyya, D. K. (2012). *A DDoS attack detection mechanism based on protocol specific traffic features*. Paper presented at the Proceedings of the Second International Conference on Computational Science, Engineering and Information Technology, Coimbatore UNK, India. https://doi.org/10.1145/2393216.2393249

[146] Huang, K., Siegel, M., & Madnick, S. Systematically Understanding the Cyber Attack Business: A Survey. *ACM Comput. Surv., 51*(4), Article 70. (2018)

[147] Sajeev, A., & Rajamani, H. S. (2020, 3-5 Nov. 2020). *Cyber-Attacks on Smart Home Energy Management Systems under Aggregators*. Paper presented at the 2020 International Conference on Communications, Computing, Cybersecurity, and Informatics (CCCI).

[148] Khouzani, M. H. R., Sarkar, S., & Altman, E. Maximum damage malware attack in mobile wireless networks. *IEEE/ACM Trans. Netw., 20*(5), 1347–1360. (2012)

[149] Fernandes, E., Rahmati, A., Jung, J., & Prakash, A. Security Implications of Permission Models in Smart-Home Application Frameworks. *IEEE Security & Privacy, 15*(2), 24-30. (2017)

[150] Rahman, M., Carbunar, B., & Banik, M. Fit and vulnerable: Attacks and defenses for a health monitoring device. *arXiv preprint arXiv:1304.5672*. (2013)

[151] Leitão, R. (2019). *Anticipating Smart Home Security and Privacy Threats with Survivors of Intimate Partner Abuse*. Paper presented at the Proceedings of the 2019 on Designing Interactive Systems Conference, San Diego, CA, USA. https://doi.org/10.1145/3322276.3322366

[152] Bojinov, H., Bursztein, E., Lovett, E., & Boneh, D. Embedded management interfaces: Emerging massive insecurity. *BlackHat USA, 1*(8), 14. (2009)

[153] Lee, M., Lee, K., Shim, J., Cho, S.-j., & Choi, J. (2016). *Security threat on wearable services: Empirical study using a commercial smartband.* Paper presented at the 2016 IEEE International Conference on Consumer Electronics-Asia (ICCE-Asia).

[154] Min, B., & Varadharajan, V. (2015). *Design and evaluation of feature distributed malware attacks against the Internet of Things (IoT).* Paper presented at the 2015 20th International Conference on Engineering of Complex Computer Systems (ICECCS).

[155] Tekeoglu, A., & Tosun, A. Ş. (2015). *A closer look into privacy and security of Chromecast multimedia cloud communications.* Paper presented at the 2015 IEEE Conference on Computer Communications Workshops (INFOCOM WKSHPS).

[156] Tekeoglu, A., & Tosun, A. S. (2015). *Investigating security and privacy of a cloud-based wireless IP camera: NetCam.* Paper presented at the 2015 24th International Conference on Computer Communication and Networks (ICCCN).

[157] Lotfy, K., & Hale, M. L. (2016). *Assessing pairing and data exchange mechanism security in the wearable Internet of Things.* Paper presented at the 2016 IEEE International Conference on Mobile Services (MS).

[158] Tzezana, R. Scenarios for crime and terrorist attacks using the internet of things. *European Journal of Futures Research, 4*(1), 18. (2016)

[159] Alghamdi, F. A., & Alghamdi, A. A. (2023, 6-8 Dec. 2023). *Navigating the Privacy and Cybersecurity Risks of Smart Homes.* Paper presented at the 2023 24th International Arab Conference on Information Technology (ACIT).

[160] Azzedin, F., & Al-Hejri, I. (2024). *Internet-of-Things Data Security: Challenges, Countermeasures and Opportunities*. Paper presented at the Proceedings of the 7th International Conference on Future Networks and Distributed Systems, Dubai, United Arab Emirates. https://doi.org/10.1145/3644713.3644808

[161] Bugeja, J., Jacobsson, A., & Davidsson, P. (2016). *On privacy and security challenges in smart connected homes.* Paper presented at the 2016 European Intelligence and Security Informatics Conference (EISIC).

[162] DeMarinis, N., & Fonseca, R. (2017). *Toward usable network traffic policies for IoT devices in consumer networks.* Paper presented at the Proceedings of the 2017 Workshop on Internet of Things Security and Privacy.

[163] Ahmad, W., Sunshine, J., Kaestner, C., & Wynne, A. (2015). *Enforcing fine-grained security and privacy policies in an ecosystem within an ecosystem.* Paper presented at the Proceedings of the 3rd International Workshop on Mobile Development Lifecycle.

[164] Hoang, N. P., & Pishva, D. (2015). *A TOR-based anonymous communication approach to secure smart home appliances.* Paper presented at the 2015

17th International Conference on Advanced Communication Technology (ICACT).
[165] Anwar, M. N., Nazir, M., & Mustafa, K. (2017, 15-16 Sept. 2017). *Security threats taxonomy: Smart-home perspective.* Paper presented at the 2017 3rd International Conference on Advances in Computing,Communication & Automation (ICACCA) (Fall).
[166] Jakobi, T., Alizadeh, F., Marburger, M., & Stevens, G. (2021). *A Consumer Perspective on Privacy Risk Awareness of Connected Car Data Use*. Paper presented at the Proceedings of Mensch und Computer 2021, Ingolstadt, Germany. https://doi.org/10.1145/3473856.3473891
[167] Winter, J. S. Citizen perspectives on the customization/privacy paradox related to smart meter implementation. *International Journal of Technoethics (IJT), 6*(1), 45-59. (2015)
[168] Copos, B., Levitt, K., Bishop, M., & Rowe, J. (2016). *Is anybody home? inferring activity from smart home network traffic.* Paper presented at the 2016 IEEE Security and Privacy Workshops (SPW).
[169] Fafoutis, X., Marchegiani, L., Papadopoulos, G. Z., Piechocki, R., Tryfonas, T., & Oikonomou, G. Privacy leakage of physical activity levels in wireless embedded wearable systems. *IEEE Signal Processing Letters, 24*(2), 136-140. (2016)
[170] He, J., Xiao, Q., He, P., & Pathan, M. S. An adaptive privacy protection method for smart home environments using supervised learning. *Future Internet, 9*(1), 7. (2017)
[171] Park, H., Basaran, C., Park, T., & Son, S. H. Energy-efficient privacy protection for smart home environments using behavioral semantics. *SENSORS, 14*(9), 16235-16257. (2014)
[172] Reichherzer, T., Timm, M., Earley, N., Reyes, N., & Kumar, V. (2017). *Using machine learning techniques to track individuals & their fitness activities.* Paper presented at the CATA 2017.
[173] Sanchez, I., Satta, R., Fovino, I. N., et al. (2014). *Privacy leakages in smart home wireless technologies.* Paper presented at the 2014 international carnahan conference on security technology (ICCST).
[174] Snader, R., Kravets, R., & Harris III, A. F. (2016). *Cryptocop: Lightweight, energy-efficient encryption and privacy for wearable devices.* Paper presented at the Proceedings of the 2016 workshop on wearable systems and applications.
[175] Srinivasan, V., Stankovic, J., & Whitehouse, K. A fingerprint and timing-based snooping attack on residential sensor systems. *ACM SIGBED Review, 5*(1), 1-2. (2008)
[176] Srinivasan, V., Stankovic, J., & Whitehouse, K. (2008). *Protecting your daily in-home activity information from a wireless snooping attack.* Paper presented at the Proceedings of the 10th international conference on Ubiquitous computing.
[177] Chen, D., Kalra, S., Irwin, D., Shenoy, P., & Albrecht, J. Preventing occupancy detection from smart meters. *IEEE Transactions on Smart Grid, 6*(5), 2426-2434. (2015)
[178] Schurgot, M. R., Shinberg, D. A., & Greenwald, L. G. (2015). *Experiments with security and privacy in IoT networks.* Paper presented at the 2015 IEEE 16th International Symposium on a World of Wireless, Mobile and Multimedia Networks (WoWMoM).

[179] Anand, S. A., & Saxena, N. (2016). *Vibreaker: Securing vibrational pairing with deliberate acoustic noise.* Paper presented at the Proceedings of the 9th ACM Conference on Security & Privacy in Wireless and Mobile Networks.

[180] Yoshigoe, K., Dai, W., Abramson, M., & Jacobs, A. (2015). *Overcoming invasion of privacy in smart home environment with synthetic packet injection.* Paper presented at the 2015 TRON Symposium (TRONSHOW).

[181] Das, A. K., Pathak, P. H., Chuah, C.-N., & Mohapatra, P. (2016). *Uncovering privacy leakage in BLE network traffic of wearable fitness trackers.* Paper presented at the Proceedings of the 17th international workshop on mobile computing systems and applications.

[182] Torre, I., Koceva, F., Sanchez, O. R., & Adorni, G. (2016). *Fitness trackers and wearable devices: how to prevent inference risks?* Paper presented at the Proceedings of the 11th EAI International Conference on Body Area Networks.

[183] Aktypi, A., Nurse, J. R., & Goldsmith, M. (2017). Unwinding Ariadne's identity thread: Privacy risks with fitness trackers and online social networks *Proceedings of the 2017 on Multimedia Privacy and Security* (pp. 1-11).

[184] Amin, S. M., & Giacomoni, A. M. Smart Grid-Safe, Secure, Self-Healing. *IEEE power and energy magazine, 10*(1), 33-40. (2011)

[185] Aouini, I., & Azzouz, L. B. Smart meter: applications, security issues and challenges. (2015)

[186] Bergmann, O., Gerdes, S., Schäfer, S., Junge, F., & Bormann, C. (2012). *Secure bootstrapping of nodes in a CoAP network.* Paper presented at the 2012 IEEE Wireless Communications and Networking Conference Workshops (WCNCW).

[187] Kermani, M. M., Zhang, M., Raghunathan, A., & Jha, N. K. (2013). *Emerging frontiers in embedded security.* Paper presented at the 2013 26th international conference on VLSI design and 2013 12th international conference on embedded systems.

[188] Greensmith, J. (2015). *Securing the internet of things with responsive artificial immune systems.* Paper presented at the Proceedings of the 2015 annual conference on genetic and evolutionary computation.

[189] Brauchli, A., & Li, D. (2015). *A solution based analysis of attack vectors on smart home systems.* Paper presented at the 2015 International Conference on Cyber Security of Smart Cities, Industrial Control System and Communications (SSIC).

[190] Ho, G., Leung, D., Mishra, P., Hosseini, A., Song, D., & Wagner, D. (2016). *Smart locks: Lessons for securing commodity internet of things devices.* Paper presented at the Proceedings of the 11th ACM on Asia conference on computer and communications security.

[191] Agadakos, I., Chen, C.-Y., Campanelli, M., et al. (2017). *Jumping the air gap: Modeling cyber-physical attack paths in the Internet-of-Things.* Paper presented at the Proceedings of the 2017 workshop on cyber-physical systems security and privacy.

[192] Oluwafemi, T., Kohno, T., Gupta, S., & Patel, S. (2013). *Experimental security analyses of {Non-Networked} compact fluorescent lamps: A case study of home automation security.* Paper presented at the LASER 2013 (LASER 2013).

[193] Denning, T., Matuszek, C., Koscher, K., Smith, J. R., & Kohno, T. (2009). *A spotlight on security and privacy risks with future household robots: attacks*

*and lessons.* Paper presented at the Proceedings of the 11th international conference on Ubiquitous computing.

[194] Ganguly, P., Poddar, S., Dutta, S., & Nasipuri, M. (2016). *Analysis of the security anomalies in the smart metering infrastructure and its impact on energy profiling and measurement.* Paper presented at the 2016 5th International Conference on Smart Cities and Green ICT Systems (SMARTGREENS).

[195] Obermaier, J., & Hutle, M. (2016). *Analyzing the security and privacy of cloud-based video surveillance systems.* Paper presented at the Proceedings of the 2nd ACM international workshop on IoT privacy, trust, and security.

[196] Bachy, Y., Basse, F., Nicomette, V., et al. (2015). *Smart-TV security analysis: practical experiments.* Paper presented at the 2015 45th Annual IEEE/IFIP International Conference on Dependable Systems and Networks.

[197] Feng, X., Ye, M., Swaminathan, V., & Wei, S. (2017). *Towards the security of motion detection-based video surveillance on IoT devices.* Paper presented at the Proceedings of the on Thematic Workshops of ACM Multimedia 2017.

[198] Xu, H., Sgandurra, D., Mayes, K., Li, P., & Wang, R. (2017). *Analysing the resilience of the internet of things against physical and proximity attacks.* Paper presented at the Security, Privacy, and Anonymity in Computation, Communication, and Storage: SpaCCS 2017 International Workshops, Guangzhou, China, December 12-15, 2017, Proceedings 10.

[199] Gu, J., Wang, J., Yu, Z., & Shen, K. Traffic-Based Side-Channel Attack in Video Streaming. *IEEE/ACM Trans. Netw., 27*(3), 972–985. (2019)

[200] Kumar, P., Gurtov, A., Iinatti, J., Ylianttila, M., & Sain, M. Lightweight and secure session-key establishment scheme in smart home environments. *IEEE Sensors Journal, 16*(1), 254-264. (2015)

[201] Coppolino, L., D'Alessandro, V., D'Antonio, S., Levy, L., & Romano, L. (2015). *My smart home is under attack.* Paper presented at the 2015 IEEE 18th International Conference on Computational Science and Engineering.

[202] Arabo, A. Cyber security challenges within the connected home ecosystem futures. *Procedia Computer Science, 61*, 227-232. (2015)

[203] Mosenia, A., Sur-Kolay, S., Raghunathan, A., & Jha, N. K. DISASTER: dedicated intelligent security attacks on sensor-triggered emergency responses. *IEEE Transactions on Multi-Scale Computing Systems, 3*(4), 255-268. (2017)

[204] Murillo, M. On Vulnerabilities of I o T-Based Consumer-Oriented Closed-Loop Control Automation Systems. *Cyber-Assurance for the Internet of Things*, 187-208. (2016)

[205] Vigo, R., Yüksel, E., & Ramli, C. D. P. K. (2012). *Smart grid security a smart meter-centric perspective.* Paper presented at the 2012 20th Telecommunications forum (TELFOR).

[206] Ul Rehman, S., & Manickam, S. A study of smart home environment and its security threats. *International Journal of Reliability, Quality and Safety Engineering, 23*(03), 1640005. (2016)

[207] Vemi, S. G., & Panchev, C. (2015). *Vulnerability testing of wireless access points using unmanned aerial vehicles (uav).* Paper presented at the Proceedings of the European Conference on e-Learning.

[208] Al Delail, B., & Yeun, C. Y. (2015). *Recent advances of smart glass application security and privacy.* Paper presented at the 2015 10th International Conference for Internet Technology and Secured Transactions (ICITST).

[209] Jacobsson, A., Boldt, M., & Carlsson, B. A risk analysis of a smart home automation system. *Future Generation Computer Systems, 56*, 719-733. (2016)
[210] Lyu, M., Sherratt, D., Sivanathan, A., Gharakheili, H. H., Radford, A., & Sivaraman, V. (2017). *Quantifying the reflective DDoS attack capability of household IoT devices.* Paper presented at the Proceedings of the 10th ACM Conference on Security and Privacy in Wireless and Mobile Networks.
[211] Tweneboah-Koduah, S., Skouby, K. E., & Tadayoni, R. Cyber security threats to IoT applications and service domains. *Wireless Personal Communications, 95*, 169-185. (2017)
[212] Abraham, O. A., Ochiai, H., Hossain, M. D., Taenaka, Y., & Kadobayashi, Y. (2023, 26-28 April 2023). *Electricity Theft Detection for Smart Homes with Knowledge-Based Synthetic Attack Data.* Paper presented at the 2023 IEEE 19th International Conference on Factory Communication Systems (WFCS).
[213] Abraham, O. A., Ochiai, H., Hossain, M. D., Taenaka, Y., & Kadobayashi, Y. Electricity Theft Detection for Smart Homes: Harnessing the Power of Machine Learning With Real and Synthetic Attacks. *IEEE Access, 12*, 26023-26045. (2024)
[214] Lo, C.-H., & Ansari, N. CONSUMER: A novel hybrid intrusion detection system for distribution networks in smart grid. *IEEE Transactions on Emerging Topics in Computing, 1*(1), 33-44. (2013)
[215] Liu, Y., Hu, S., Wu, J., et al. (2015). *Impact assessment of net metering on smart home cyberattack detection.* Paper presented at the Proceedings of the 52nd Annual Design Automation Conference.
[216] Liu, Y., & Hu, S. Cyberthreat analysis and detection for energy theft in social networking of smart homes. *IEEE Transactions on Computational Social Systems, 2*(4), 148-158. (2015)
[217] Liu, Y., Hu, S., & Zomaya, A. Y. The Hierarchical Smart Home Cyberattack Detection Considering Power Overloading and Frequency Disturbance. *IEEE Transactions on Industrial Informatics, 12*(5), 1973-1983. (2016)
[218] Liu, Y., Hu, S., & Ho, T. Y. Leveraging Strategic Detection Techniques for Smart Home Pricing Cyberattacks. *IEEE Transactions on Dependable and Secure Computing, 13*(2), 220-235. (2016)
[219] Liu, J., & Sun, W. Smart attacks against intelligent wearables in people-centric internet of things. *IEEE Communications Magazine, 54*(12), 44-49. (2016)
[220] Liu, Y., Zhou, Y., & Hu, S. Combating Coordinated Pricing Cyberattack and Energy Theft in Smart Home Cyber-Physical Systems. *IEEE Transactions on Computer-Aided Design of Integrated Circuits and Systems, 37*(3), 573-586. (2018)
[221] Tushir, B., Dalal, Y., Dezfouli, B., & Liu, Y. A Quantitative Study of DDoS and E-DDoS Attacks on WiFi Smart Home Devices. *IEEE Internet of Things Journal, 8*(8), 6282-6292. (2021)
[222] Bugeja, J., Jacobsson, A., & Davidsson, P. (2017). *An analysis of malicious threat agents for the smart connected home.* Paper presented at the 2017 IEEE international conference on pervasive computing and communications workshops (PerCom Workshops).
[223] Chen, Y., & Luo, B. (2012). *S2a: secure smart household appliances.* Paper presented at the Proceedings of the second ACM conference on Data and Application Security and Privacy.

[224] Kang, W. M., Moon, S. Y., & Park, J. H. An enhanced security framework for home appliances in smart home. *Human-centric Computing and Information Sciences, 7*, 1-12. (2017)

[225] Komninos, N., Philippou, E., & Pitsillides, A. Survey in smart grid and smart home security: Issues, challenges and countermeasures. *IEEE Communications Surveys & Tutorials, 16*(4), 1933-1954. (2014)

[226] Fernandes, E., Jung, J., & Prakash, A. (2016). *Security analysis of emerging smart home applications.* Paper presented at the 2016 IEEE symposium on security and privacy (SP).

[227] Hodges, D. Cyber-enabled burglary of smart homes. *COMPUTERS & SECURITY, 110*. (2021)

[228] Tzezana, R. High-probability and wild-card scenarios for future crimes and terror attacks using the Internet of Things. *foresight, 19*(1), 1-14. (2017)

[229] Elkhail, A. A., Refat, R. U. D., Habre, R., Hafeez, A., Bacha, A., & Malik, H. Vehicle Security: A Survey of Security Issues and Vulnerabilities, Malware Attacks and Defenses. *IEEE Access, 9*, 162401-162437. (2021)

[230] Anthi, E., Williams, L., Javed, A., & Burnap, P. Hardening machine learning denial of service (DoS) defences against adversarial attacks in IoT smart home networks. *COMPUTERS & SECURITY, 108*, 1. (2021)

[231] Jayasinghe, K., & Poravi, G. (2020). *A Survey of Attack Instances of Cryptojacking Targeting Cloud Infrastructure*. Paper presented at the Proceedings of the 2020 2nd Asia Pacific Information Technology Conference, Bali Island, Indonesia. https://doi.org/10.1145/3379310.3379323

[232] Alizadeh, F., Stevens, G., Jakobi, T., & Krüger, J. Catch Me if You Can : "Delaying" as a Social Engineering Technique in the Post-Attack Phase. *Proc. ACM Hum.-Comput. Interact., 7*(CSCW1), Article 32. (2023)